\documentclass{rsproca}

\newenvironment{remark}{ \noindent \textbf{Remark: }}

\usepackage{subfigure}

\begin{document}

\title{On a consistent finite-strain plate theory based on 3-D energy principle}

\author{ Hui-Hui Dai$^{1}$, Zilong Song$^{1}$ }

\address{$^{1}$ Department of Mathematics, City University of Hong Kong,
83 Tat Chee Avenue, Kowloon Tong, Hong Kong}

\subject{Solid Mechanics, Applied mathematics}


\corres{Hui-Hui Dai\\
\email{mahhdai@cityu.edu.hk}\\
Zilong Song\\
\email{buctsongzilong@163.com}}

\begin{abstract}
This paper derives a finite-strain plate theory consistent with the principle of stationary three-dimensional (3-D) potential energy under general loadings with a third-order error. Staring from the 3-D nonlinear elasticity (with both geometrical and material nonlinearity) and by a series expansion, we deduce a vector plate equation with three unknowns, which exhibits the local force-balance structure.  The success relies on using the 3-D field equations and bottom traction condition to derive exact recursion relations for the coefficients. Associated weak formulations are considered, leading to a 2-D virtual work principle. An alternative approach based on a 2-D truncated energy is also provided, which is less consistent than the first plate theory but has the advantage of the existence of a 2-D energy function. As an example, we consider the pure bending problem of a hyperelastic block. The comparison between the analytical plate solution and available exact one shows that the plate theory gives second-order correct results. Comparing with existing plate theories, it appears that the present one has a number of advantages, including  the consistency, order of correctness, generality of the loadings, applicability to finite-strain problems and no involvement of unphysical quantities.
\end{abstract}


\begin{fmtext}
\section{Introduction}
Plates are very important engineering structures, which have attracted
extensive research since the 19th century. Plates are defined as plane
structural elements with a small thickness (characteristic length $h$)
compared with the other two planar dimensions. Plate theories attempt
to reduce the three-dimensional (3-D) elasticity theory to a 2-D
approximate one defined on a surface. The literature on plate theories
is vast, including direct \cite{altenbach2010,altenbach2014} and
derived plate theories. Here, we only give a review on selected works
on derived plate theories, which are
\end{fmtext}

\maketitle


\noindent
divided into three categories.

(1) A common starting point is the series expansion of the deformed
position (or displacement) vector  in terms of thickness variable $Z$,
like
\begin{equation}
\label{eq0}
\begin{aligned}
\mathbf{x}(\mathbf{r},Z) = \sum_{j=0}^N Z^j \mathbf{x}^{(j)}
(\mathbf{r}).
\end{aligned}
\end{equation}
Early attempts on plate theories relied on {\it a priori} hypotheses,
mostly motivated by engineering intuition. The classical plate theory
(Kirchhoff-Love theory, see Kirchhoff \cite{kirchhoff1850} and Love
\cite{love1888}), which relies on three assumptions about both the
geometry and deformation, is known to be only applicable to thin
plates. Within linear elasticity, the first-order shear deformable
plate theory \cite{mindlin1951} (Mindlin-Reissner theory) relaxes one
of Kirchhoff's assumptions, and introduces additional two unknowns. As
a refinement, the third-order shear deformable plate theory
\cite{reddy2004,reddy2007} incorporates postulated cubic terms in the
two planar displacements, by assuming the way that the stresses vary
over a cross-section. The advantage is that it avoids the use of the
shear correction factor, whereas the results for $\mathbf{x}$ are
almost the same  as in Mindlin-Reissner theory (see \cite{reddy2007}).
The von K\'{a}rm\'{a}n plate theory \cite{von1910} still uses
Kirchhoff's assumptions but retains some nonlinear components for the
strain tensor (geometric nonlinearity) with an attempt to describe
large deflections for thin plates. Although widely accepted and used,
due to the hypotheses involved, one cannot expect that they are
consistent with the 3-D formulation up to the required order for
general loadings. Also, they do not expect to provide good results for
relatively thick plates and certainly cannot be used for finite-strain
problems.

(2) Another approach is also based on $(\ref{eq0})$  but with no
explicit kinematic assumptions. All the coefficients
$\mathbf{x}^{(j)}$ ($j\ge 1$) are treated as independent unknowns,
whose governing equations are derived from the 2-D variational (or
virtual work) principle by first integrating out the $Z$ variable and
conducting a truncation. Such an approach was adopted by Kienzler
\cite{kienzler2002,kienzler2012} based on  linear elasticity. By a
procedure of "pseudo-reduction" of the resulting system to certain
orders of $h$, the classical plate theory and a Reissner-type theory
\cite{reissner1945, reissner1986} were recovered. Based on nonlinear
elasticity, Meroueh \cite{Meroueh1986} used a similar approach except
with Legendre polynomials of $Z$ in $(\ref{eq0})$, and formulated a
system in terms of generalized (higher-order) stress resultants for
finite-strain problems. Also based on nonlinear elasticity, Steigmann
\cite{steigmann2007} carried out a study to construct an $O(h^3)$ 2-D
energy and finally arrived at a fourth-order system for
$\mathbf{x}^{(0)}$ by eliminating other unknowns $\mathbf{x}^{(j)}$
($j\ge 1$). The disadvantage of these theories is that the resulting
system (Euler-Lagrange equation) contains many unknowns, albeit some
may be eliminated by a further reduction. In addition, it is not easy
to interpret and propose suitable and consistent boundary conditions
on the lateral surface, since they involve generalized (high-order)
stress resultants whose physical meanings are not clear. Also, as
pointed out by Steigmann \cite{steigmann2013}, for the truncated
energy to be as accurate as possible by the standard of 3-D elasticity
theory, one needs to impose restrictions on the high-order
coefficients in $(\ref{eq0})$ from the 3-D formulation instead of
treating them as independent unknowns. This way was adopted by
Steigmann in \cite{steigmann2008, steigmann2010,steigmann2013},
leading to a significant progress towards the derivation of more
proper plate and shell models which incorporate both stretching and
bending. More specifically, the author considered the case that the
tractions were zero (or sufficiently small) on the top and bottom
surfaces, which were utilized to represent $\mathbf{x}^{(i)}$ ($i=
1,2$) in terms of $\mathbf{x} ^{(0)}$, leading to a final system with
only $\mathbf{x}^{(0)}$. The theory appears to extend Koiter's shell
theory \cite{ciarlet2005} and dictate an optimal $O(h^3)$
approximation for the 3-D potential energy. However,  as indicated by
the author in \cite{steigmann2013}, an undesirable feature is that the
relation for $\mathbf{x} ^{(1)}$ is not accurate enough, which may
cause a non-negligible $O(h^3)$ error in the energy. And, the theory
is restricted to the traction-free (or sufficiently small traction)
case.

(3) Some consistent mathematical approaches for deriving leading-order
plate theories have also been developed, which are based on certain
{\it a priori} scalings between the thickness $h$ and the deformations
(or applied loads). The method of Gamma convergence
\cite{ledret1993,ledret1995,friesecke2006} concerns with the limiting
2-D variational problem of vanishing thickness $h$, which leads to a
hierarchy of 2-D energies depending on the scalings. This method is
rigorous but so far has failed to furnish a model containing the
thickness parameter which incorporates both bending and stretching. On
the other hand, asymptotic methods \cite{ciarlet1980, erbay1997,
millet2001} aim at generating the leading-order (in $h$) 2-D
variational problem or differential system via formal expansions. In
\cite{ciarlet1980}, it was shown that von K\'{a}rm\'{a}n plate
equations could be derived formally by such an approach based on the
3-D weak formulation with prescribed orders of applied loads and
certain lateral surface boundary conditions. In
\cite{erbay1997,millet2001}, the derivations were based on the 3-D
differential formulation, in particular a hierarchy of leading-order
plate equations were derived in \cite{millet2001}. However, the main
difficulty preventing the wide applicability of these models is that
they do not furnish a single plate model for all orders of applied
loads, which is perhaps most needed in engineering problems. There are
also some restrictions on the lateral boundary conditions in the
asymptotic approaches. And, from the leading-order equations, it is
difficult to examine the effects of the thickness.

Strictly speaking, a plate theory can be said to be consistent if the
approximations ensure that either the 3-D differential formulation
(including field equations and boundary conditions) or the 3-D weak
formulation (by the energy principle) are satisfied to the required
order of $h$. In our view,  a {\it good} plate theory should be
consistent with the 3-D formulation beyond the leading order with no
special restrictions on applied loads, and further should furnish
a single model with no unphysical quantities, leading to applicability
from thin plates to relatively thick plates for a variety of
deformations from small bending and stretching to finite strains.
Despite more than 150 years research and significant advancements (see
the above review for selected works), it appears that such a good
plate theory has not been established. In this paper, an attempt is
made in this direction.

Our starting point is also $(\ref{eq0})$ (with $N=4$). With this
expansion and based on 3-D nonlinear elasticity, we intend to derive
(without special restrictions on applied loads) a 2-D plate system
consistent with the 3-D stationary potential energy principle (or weak
formulation) with an error of $O(h^4)$, with an aim of producing
$O(h^2)$-correct results for all the displacement (position) vector,
strain tensor and stress tensor. We mention some key points for our
success. First, rather than obtaining a truncated 2-D energy by
integration, we deduce the main results by directly approximating the
exact 3-D field equations and conditions on the top and bottom
surfaces. This was the way adopted before (with only third-order
material nonlinearity) in \cite{dai2006,dai2012b,dai2013} for deriving
rod-like theories and in \cite{dai2012} for deriving a plate model
involving only stretching. Specifically, those 3-D differential
relations are kept to the desired order in a pointwise manner, such
that the corresponding terms in the variation of the 3-D potential
energy are $O(h^4)$. A finding is that the exact recursive relations
between $\mathbf{x}^{(i)} (i=2,3,4)$ and $\mathbf{x}^{(i)} (i=0,1)$
can be obtained by solving {\it linear algebraic equations}. Second,
we adopt an expansion about the bottom surface, as did in
\cite{dai2014} for a tube buckling problem. This enables us to derive
an exact algebraic relation between $\mathbf{x}^{(1)}$ and $\nabla
\mathbf{x}^{(0)}$, which avoids the non-negligible error in
\cite{steigmann2013}. These relations show the dependence among
$\mathbf{x}^{(i)} (i=0,1,2,3,4)$,  and thus it may not be proper to
treat them as independent unknowns as did in some works. As a result,
the final vector plate equation only contains three knowns
$\mathbf{x}^{(0)}$. Proper plate boundary conditions are then
introduced to make the two edge-integration terms in the 3-D energy
principle be $O(h^4)$.

Associated weak formulations for the derived differential plate system
are also provided, which show that the latter obeys a 2-D virtual work principle 
and lead to the natural boundary conditions. By relaxing
the consistent criterion a little, we also construct a truncated 2-D
strain energy as did in \cite{steigmann2013}. The differences are that it is not restricted to the
traction-free case and there is no non-negligible error. The advantage
of this approach is that the plate problem can be solved by an energy
minimization, although the 3-D weak formulation is not satisfied
exactly to $O(h^4)$.  To examine the vadility of our consistent plate
theory, we consider the pure finite-bending problem of a rectangular
hyperelastic block, for which the exact solutions are available. The
comparison between the analytical plate solution and the exact one
supports our claims that the plate theory can provide $O(h^2)$-correct
results for the displacement vector, strain tensor and stress tensor.
It appears that no existing plate theories have been demonstrated to
produce results correct to this order. Finally, we give some concluding
remarks, including a summary of nice features of this plate theory.

\section{The 3-D energy principle and field equations}

We consider a homogeneous thin plate of constant thickness composed of
a hyperelastic material. A material point in the reference
configuration $\kappa=\Omega\times [0,2h]$ is denoted by $\mathbf{X} =
(\mathbf{r},Z)$, where the thickness $2 h$ of the plate is small
compared with the planar dimensions of the bottom surface $\Omega$.
The deformed position in the current configuration $\kappa_t$ is
denoted by $\mathbf{x}$. In this section, we first recall the 3-D
formulation and then introduce the consistency criterion for a plate
theory.

For a plate structure, the deformation gradient is represented as
\begin{equation}
\label{eq6}
\begin{aligned}
\mathbf{F}= \frac{\partial \mathbf{x}}{\partial \mathbf{X}} =
\frac{\partial \mathbf{x}}{\partial \mathbf{r}} + \frac{\partial
\mathbf{x}}{\partial Z} \otimes \mathbf{k} = \nabla \mathbf{x} +
\frac{\partial \mathbf{x}}{\partial Z} \otimes \mathbf{k},
\end{aligned}
\end{equation}
where $\nabla$ is the in-plane two-dimensional gradient, and
$\mathbf{k}$ is the unit normal to the reference bottom surface
$\Omega$. For a hyperelastic material, the nominal stress $\mathbf{S}$
can be obtained through the strain energy function  $\Phi(\mathbf{F})$
by $\mathbf{S}(\mathbf{F}) = \frac{\partial \Phi }{\partial
\mathbf{F}}.$ The associated first and second order elastic moduli are
defined by
\begin{equation}
\label{eq1_2}
\begin{aligned}
\mathcal{A}^{1}(\mathbf{F}) = \frac{\partial^2 \Phi }{\partial
\mathbf{F}\partial \mathbf{F}}~~ (\mathcal{A}_{ijkl}^{1} =
\frac{\partial^2 \Phi }{\partial {F}_{ji}\partial {F}_{lk}}),\quad
\mathcal{A}^{2}(\mathbf{F}) = \frac{\partial^3 \Phi }{\partial
\mathbf{F}\partial \mathbf{F}\partial \mathbf{F}}.
\end{aligned}
\end{equation}
Here and in the sequel we adopt the convention that Latin indices run
from 1 to 3 whereas Greek indices run from 1 to 2, and the index after
the comma indicates differentiation (e.g. $(\nabla
\mathbf{x})_{i\alpha}=\mathbf{x}_{i,\alpha}$ in $(\ref{eq6})$). It is
assumed that the strain energy function for the deformations concerned
satisfies the strong-ellipticity condition
\begin{equation}
\label{eq1_4}
\begin{aligned}
\mathbf{a} \otimes \mathbf{b} : \mathcal{A}^{1}(\mathbf{F})[\mathbf{a}
\otimes \mathbf{b}] >0,\quad \mathrm{for \ all}\ \mathbf{a} \otimes
\mathbf{b} \ne 0,
\end{aligned}
\end{equation}
where the colon means a scalar tensor product $\mathbf{A}: \mathbf{B}=
\mathrm{tr} (\mathbf{A} \mathbf{B}^T)$ and the square bracket after a
modulus tensor means the operation:
$(\mathcal{A}^{1}[\mathbf{A}])_{ij} = \mathcal{A}_{ijkl}^{1} {A}_{lk}$
in rectangular cartesian coordinates.

For the case of dead-loading and in the absence of body forces, the
3-D potential energy $E$ is given by
\begin{equation}
\label{eq1}
\begin{aligned}
& E= \bar{\Phi} -\bar{V}, \qquad  \bar{\Phi} = \int_{\Omega}
\int_{0}^{2 h} \Phi(\mathbf{F}) \mathrm{dZ} \mathrm{d\mathbf{r}},\\
&\bar{V}= \int_{\Omega} \mathbf{q}^{-} (\mathbf{r}) \cdot
\mathbf{x}(\mathbf{r},0) -\mathbf{q}^{+}(\mathbf{r})  \cdot
\mathbf{x}(\mathbf{r},2h) \ \mathrm{d\mathbf{r}} + \int_{\partial
\Omega_q} \int_{0}^{2h} \mathbf{q}(s,Z) \cdot
\mathbf{x}(s,Z)\mathrm{dZds},
\end{aligned}
\end{equation}
where $\bar{V}$ is the load potential, $\mathbf{q}^{\pm}$ are the
applied tractions on the top and bottom surfaces, and $\mathbf{q}$ is
the applied traction on the lateral surface (the edge). Here the
boundary $\partial \Omega$ is divided into two parts, the position
boundary $\partial \Omega_0$ and the traction boundary $\partial
\Omega_q$. The principle of stationary potential energy requires the
first variation of $E$ to be zero, which leads to
\begin{equation}
\label{eq2}
\begin{aligned}
\delta E =& - \int_{\Omega} \int_{0}^{2 h} \mathrm{Div}
\mathbf{S}\cdot \delta \mathbf{x} \mathrm{dZ d\mathbf{r}}
-\int_{\Omega} \left(\left.\mathbf{S}^T \mathbf{k}\right|_{Z=0} +
\mathbf{q}^{-}\right) \cdot \delta \mathbf{x}(\mathbf{r},0) \mathrm{d\mathbf{r}}\\
&+ \int_{\Omega} \left(\left. \mathbf{S}^T \mathbf{k}\right|_{Z=2h} -
\mathbf{q}^{+}\right) \cdot \delta \mathbf{x}(\mathbf{r},2 h)
\mathrm{d\mathbf{r}} + \int_{\partial \Omega_0} \int_{0}^{2h}
\mathbf{S}^T \mathbf{N}
\cdot \delta \mathbf{x}(s,Z)\mathrm{dZds}\\
&+\int_{\partial \Omega_q} \int_{0}^{2h} \left(\mathbf{S}^T \mathbf{N}
- \mathbf{q}\right) \cdot \delta \mathbf{x}(s,Z)\mathrm{dZds}=0,
\end{aligned}
\end{equation}
where $\mathbf{N}$ is the unit outward normal to the lateral surface.
This is a 3-D weak formulation of the problem. Then the 3-D field
equations together with boundary conditions are
\begin{equation}
\label{eq3}
\begin{aligned}
&\mathrm{Div} \mathbf{S} =0,\quad \mathrm{in} \quad \Omega\times [0,2h],\\
&\left.\mathbf{S}^T \mathbf{k}\right|_{Z=0} = -\mathbf{q}^{-},\quad
\left.\mathbf{S}^T \mathbf{k}\right|_{Z=2h} = \mathbf{q}^{+},
\quad \mathrm{in} \quad \Omega\\
& \mathbf{x} = \mathbf{b} (s,Z), \quad \mathrm{on} \quad \partial
\Omega_0 \times [0,2h], \\
&\mathbf{S}^T \mathbf{N} = \mathbf{q}(s,Z),\quad \mathrm{on} \quad
\partial \Omega_q \times [0,2h],
\end{aligned}
\end{equation}
where $\mathbf{b}$ is the prescribed position on the boundary
$\partial \Omega_0$.

For a consistent plate theory, one needs to make approximations to
eliminate the $Z$ variable, which should agree with the principle of
stationary 3-D potential energy (i.e. the 3-D weak formulation) to
certain order. Thus, the consistency criterion is that for {\it all}
loadings (without {\it priori} restrictions on
$\mathbf{q}^{\pm},\mathbf{q}(s,Z)$ and $\mathbf{b} (s,Z)$ except some
smooth requirement) each term in $\delta E$ should be either zero or a
required asymptotic order (say, $O(h^4)$) {\it separately} for the
plate approximation. It should be noted that $\delta \mathbf{x}$ is
not correlated in each of these terms. Even a plate model can make
$\delta E$  be of the required asymptotic order as a whole, it does
not guarantee each term to be of that order, and thus it could still
be inconsistent with the 3-D weak formulation. As far as the authors
are aware of, no existing plate theories satisfy the above criterion
up to $O(h^4)$. The main purpose of the present paper is to provide
such a consistent plate theory.

\section{The 2-D vector plate equation}

The starting point of our derivation of a consistent plate theory is a
series expansion of the current position vector, which is employed
with the previously defined consistency criterion in mind. First, we
consider the corresponding expansions of the deformation gradient and
nominal stress and make some key observations, which are essential for
the success of our procedure. In the sequel, without loss of generality, it is understood that all the spatial variables and position/displacement vectors are scaled by the typical length of the in-plane surface and all the stresses, energies and applied tractions are scaled by a typical stress magnitude. Then, in particular, $2h$ means the thickness ratio.

\subsection{Expansions}

In order to make approximations to obtain a 2-D formulation, a
plausible way is to take the advantage of the thinness of the plate by
taking series in $Z$. Suppose that the current position vector
$\mathbf{x}(\mathbf{X})$ is a $C^5$ function in $Z$, then for any
$0\le Z\le 2h$ we can expand it about $Z=0$ (the bottom surface) as
\begin{equation}
\label{eq4}
\begin{aligned}
\mathbf{x}(\mathbf{X}) = \mathbf{x}^{(0)}(\mathbf{r}) + Z
\mathbf{x}^{(1)}(\mathbf{r}) + \frac{1}{2} Z^2 \mathbf{x}^{(2)}
(\mathbf{r}) +  \frac{1}{6} Z^3 \mathbf{x}^{(3)} (\mathbf{r}) +
\frac{1}{24} Z^4 \mathbf{x}^{(4)}(\mathbf{r}) +Z^5
\mathbf{x}^{(5)}(\mathbf{r}, Z^*),
\end{aligned}
\end{equation}
where $0<Z^*<2h$ and the superscript $n$ denotes the $nth$-order
derivative and $\mathbf{x}^{(n)}(\mathbf{r}) =\left.{\frac{\partial^n
\mathbf{x}}{\partial Z^n}}\right|_{Z=0}(n=1,2,3,4).$ Accordingly, the
deformation gradient has a similar expansion
\begin{equation}
\label{eq7}
\begin{aligned}
\mathbf{F} = \mathbf{F}^{(0)}(\mathbf{r}) + Z
\mathbf{F}^{(1)}(\mathbf{r}) + \frac{1}{2} Z^2 \mathbf{F}^{(2)}
(\mathbf{r}) +  \frac{1}{6} Z^3 \mathbf{F}^{(3)} (\mathbf{r}) +O(Z^4),
\end{aligned}
\end{equation}
where $ \mathbf{F}^{(n)}$ is defined in the same way as
$\mathbf{x}^{(n)}$. Substituting $(\ref{eq4})$  into $(\ref{eq6})$ and
comparing with $(\ref{eq7})$, we obtain the relations
\begin{equation}
\label{eq8}
\begin{aligned}
\mathbf{F}^{(n)} = \nabla \mathbf{x}^{(n)} + \mathbf{x}^{(n+1)}
\otimes \mathbf{k}, \quad n=0,1,2,3.
\end{aligned}
\end{equation}
An observation is that the dependence of $\mathbf{F}^{(n)}$ on
$\mathbf{x}^{(n+1)}$ is linearly algebraic. The strain energy $\Phi$
is also assumed to belong to $C^5$ in its arguments, then the nominal
stress $\mathbf{S}$ can be expanded as
\begin{equation}
\label{eq9}
\begin{aligned}
\mathbf{S}(\mathbf{F}) = \mathbf{S}^{(0)}(\mathbf{r}) + Z
\mathbf{S}^{(1)}(\mathbf{r}) + \frac{1}{2} Z^2 \mathbf{S}^{(2)}
(\mathbf{r}) +  \frac{1}{6} Z^3 \mathbf{S}^{(3)} (\mathbf{r}) +O(Z^4).
\end{aligned}
\end{equation}
By the chain rule, the left-hand side can also be expanded in series
of $Z$ by virtue of $(\ref{eq1_2},\ref{eq7})$. Comparing two sides
leads to
\begin{equation}
\label{eq10} \mathbf{S}^{(0)}= \mathbf{S} (\mathbf{F}^{(0)}), \quad
\mathbf{S}^{(1)} = {\mathcal{A}}^1
(\mathbf{F}^{(0)})[\mathbf{F}^{(1)}],\quad \mathbf{S}^{(2)} =
{\mathcal{A}}^1 (\mathbf{F}^{(0)})[\mathbf{F}^{(2)}] + {\mathcal{A}}^2
(\mathbf{F}^{(0)})[\mathbf{F}^{(1)},\mathbf{F}^{(1)}],
\end{equation}
and in component form the last two are
\begin{equation}
\label{eq10_1}
\begin{aligned}
{S}_{ij}^{(1)} = {\mathcal{A}}_{ijkl}^1{F}_{lk}^{(1)},\quad
{S}_{ij}^{(2)} = {\mathcal{A}}_{ijkl}^1{F}_{lk}^{(2)} +
{\mathcal{A}}_{ijklmn}^2 {F}_{lk}^{(1)}{F}_{nm}^{(1)},
\end{aligned}
\end{equation}
where the argument $\mathbf{F}^{(0)}$ is omitted in ${\mathcal{A}}
^{i}$ ($i=1,2$). The relation $(\ref{eq10})$ between the components of
stress and deformation gradient is helpful for clarifying the
dependence and the sequel derivation. In practice, once the strain
energy function is specified, one can get $\mathbf{S}^{(i)} (i=0,1,2)$
directly through an expansion of $\mathbf{S}$ without computing the
moduli. Actually, $\mathbf{S}^{(3)}$ is also needed, but it is an
intermediate quantity, whose expression is omitted. One can observe
that the dependence of $\mathbf{S}^{(i)}$ ($i=1,2$) on
$\mathbf{x}^{(i+1)}$ is linearly algebraic, which is one key of the
success.

Based on the above series expansions, we are ready to derive a 2-D
plate theory, which satisfies the consistency criterion defined
before.

\subsection{Derivation of the vector plate equation}
Due to the expansion $(\ref{eq4})$, the unknowns are five vectors
$\mathbf{x}^{(i)} (i=0,..,4)$. It appears there are a little too many
unknowns but it is necessary to obtain $O(h^2)$ correct results as we
shall show. The first issue is whether a closed-system for them can be
obtained up to the proper order. The second issue is whether it is
possible to eliminate most unknowns (one wants to avoid a consistent
but too complicated plate theory). In this subsection, we shall
address both issues.

First of all, we substitute $(\ref{eq9})$ into the bottom traction
condition $(\ref{eq3})_2$ to obtain
\begin{equation}
\label{eqX1} (\mathbf{S}^{(0)})^T \mathbf{k} = - \mathbf{q}^{-}.
\end{equation}
The advantage of the expansion at the bottom is that it is an exact
equation which contains only two unknown vectors $\mathbf{x^{(0)}}$
and $\mathbf{x^{(1)}}$, and further the dependence on the latter is
algebraic (cf. $(\ref{eq10})_1$ and $(\ref{eq8})$). The
strong-ellipticity condition together with the implicit function
theorem guarantee that $\mathbf{x}^{(1)}$ can be uniquely solved in
terms of $\nabla \mathbf{x}^{(0)}$ (see \cite{steigmann2007}).

Substituting $(\ref{eq9})$ into the top traction condition
$(\ref{eq3})_3$ leads to
\begin{equation}
\label{eqX2} (\mathbf{S}^{(0)})^T \mathbf{k} +2 h (\mathbf{S}^{(1)})^T
\mathbf{k} + 2 h^2 (\mathbf{S}^{(2)})^T \mathbf{k} + \frac{4}{3} h^3
(\mathbf{S}^{(3)})^T \mathbf{k}+ O(h^4)=\mathbf{q}^{+},
\end{equation}
in which $O(h^4)$ terms are omitted. The above equation contains all
five unknown vectors, and to have a closed system one needs to have
another three vector equations which contain and only contain those
unknowns. For that purpose, we utilize the 3-D field equations
$(\ref{eq3})_1$, which can be expressed as
\begin{equation}
\label{eq11}
\begin{aligned}
\mathrm{Div} \mathbf{S} = \nabla \cdot \mathbf{S} + \frac{\partial
(\mathbf{S}^T \mathbf{k})}{\partial Z}=\mathbf{0}.
\end{aligned}
\end{equation}
After the substitution of $(\ref{eq9})$, the left-hand side becomes a
series of $Z$, and the vanishing of the coefficients of $Z^n$ leads to
\begin{equation}
\label{eq12}
\begin{aligned}
\nabla \cdot \mathbf{S}^{(n)} + (\mathbf{S}^{(n+1)})^T \mathbf{k}
=\mathbf{0}, \quad n=0,1,2.
\end{aligned}
\end{equation}
The above three vector equations, involving only $\mathbf{S}^{(k)}
(k=0,1,2,3)$, contain and only contain the above-mentioned five
unknowns vectors. So, we have a closed system. This demonstrates that
the number of coefficients in the series expansion $(\ref{eq4})$
should not be taken arbitrarily, rather it should be chosen according
to the error in the top traction condition.

We also observe that $(\ref{eq12})$ relates the higher-order
coefficient $\mathbf{S}^{(n+1)}$ (whose dependence on
$\mathbf{x}^{(n+2)}$ is linearly algebraic) to the derivatives of the
lower-order coefficient $\mathbf{S}^{(n)}$ (involving up to
$\mathbf{x}^{(n+1)}$). Therefore, this series of equations can be used
to derive the recursion relations for $\mathbf{x}^{(n)}$ ($n\geq 2$)
by solving linear algebraic equations!

We take $n=0$ as an example. Substituting $(\ref{eq8},\ref{eq10})$
into $(\ref{eq12})$, we obtain
\begin{equation}
\label{eq13}
\begin{aligned}
\mathbf{B} \mathbf{x}^{(2)} + \mathbf{f}^{(2)} =0,\quad {B}_{ij}=
\mathcal{A}_{3i3j}^1,\quad \mathbf{f}^{(2)} = ({\mathcal{A}}^1 [\nabla
\mathbf{x}^{(1)}])^T \mathbf{k} + \nabla \cdot \mathbf{S}^{(0)},
\end{aligned}
\end{equation}
where both $\mathbf{B}$ and $\mathbf{f}^{(2)}$ only involve
$\mathbf{x}^{(0)}$ and $\mathbf{x}^{(1)}$. In $(\ref{eq13})$ and
hereafter, the argument $\mathbf{F}^{(0)}$ is omitted for brevity in
$\mathbf{B}$, ${\mathcal{A}}^1$ and $\mathbf{S}^{(0)}$ (and
${\mathcal{A}}^2$ in the sequel). By the strong-ellipticity condition
$(\ref{eq1_4})$, $\mathbf{B}$ is invertible and we obtain
\begin{equation}
\label{eq14}
\begin{aligned}
\mathbf{x}^{(2)}= -\mathbf{B}^{-1} \mathbf{f}^{(2)}.
\end{aligned}
\end{equation}
Similarly, for $n=1$ we obtain the following expression of
$\mathbf{x}^{(3)}$ from $(\ref{eq12})$
\begin{equation}
\label{eq15}
\begin{aligned}
&\mathbf{x}^{(3)}= -\mathbf{B}^{-1} \mathbf{f}^{(3)},\quad
\mathbf{f}^{(3)} = ({\mathcal{A}}^1 [\nabla \mathbf{x}^{(2)}] +
{\mathcal{A}}^2 [\mathbf{F}^{(1)},\mathbf{F}^{(1)}])^T \mathbf{k} +
\nabla \cdot \mathbf{S}^{(1)}.
\end{aligned}
\end{equation}
The vector $\mathbf{x}^{(4)}$ is an intermediate quantity, whose
explicit expression is not needed, however the relation $(\ref{eq12})$
with $n=2$ as a whole will be utilized to eliminate it. Due to these
recursion relations, along with $(\ref{eqX1})$, all the higher-order
terms $\mathbf{x}^{(k)}$ ($k=1,2,3$) can be expressed by
$\mathbf{x}^{(0)}$. It should be pointed out that these are {\it
exact} relations without any approximation involved.

Finally, by subtracting $(\ref{eqX1})$ and utilizing the field
equations $(\ref{eq12})$ ($n=0,1,2$) once,  $(\ref{eqX2})$ reduces to
\begin{equation}
\label{eq17} \nabla \cdot \bar{\mathbf{S}} = -\bar{\mathbf{q}},\quad
\bar{\mathbf{q}} = (\mathbf{q}^{+} + \mathbf{q}^{-})/(2 h), \quad
\bar{\mathbf{S}} = \frac{1}{2h} \int_0^{2h} \mathbf{S} \mathrm{dZ}=
\mathbf{S}^{(0)} + h  \mathbf{S}^{(1)} + \frac{2}{3} h^2
\mathbf{S}^{(2)}+O(h^3).
\end{equation}
The first equation $(\ref{eq17})_1$ is the 2-D vector plate equation,
which involves quantities up to $\mathbf{x}^{(3)}$. After replacing
$\mathbf{x}^{(i)}$ ($i=1,2,3$), it becomes a fourth-order differential
equation for $\mathbf{x}^{(0)}$ (with an error of $O(h^3)$). Once it
is solved, an up to $O(h^2)$-correct result for $\mathbf{x}^{(0)}$ can
be obtained. Then, up to $O(h^2)$-correct results for
$\mathbf{x}^{(k)}$ ($k=1,2,3$), strains and stresses can be easily
deduced.

By definition, $\bar{\mathbf{S}}$ is the averaged stress over the
thickness, and $\bar{\mathbf{q}}$ can be regarded as the effective
body force for the plate caused by the tractions on the top and bottom
surfaces. A direct integration of the 3-D field equations
$(\ref{eq3})_1$ (also refer to $(\ref{eq11})$) over the thickness
variable followed by the use of the top and bottom boundary conditions
$(\ref{eq3})_2$ leads to precisely $\nabla \cdot \bar{\mathbf{S}} =
-\bar{\mathbf{q}}$! Therefore, the present plate equation possesses,
in a through-thickness average, the local force-balance structure in all
three directions inherited from the 3-D system.

Now, we examine the consistency according to the criterion introduced
before. For that purpose, we analyze the asymptotic orders of the
first three terms in $\delta E$ (see $(\ref{eq2})$). For the first
term, we notice that $\mathrm{Div} \mathbf{S} =\sum_{n=0}^{2}
\frac{Z^n}{n!} [\nabla \cdot \mathbf{S}^{(n)} + (\mathbf{S}^{(n+1)})^T
\mathbf{k}] + O(Z^3)$. In the derivation, $(\ref{eq12})$ with
$n=0,1,2$ were utilized to obtain the recursion relations for
eliminating $\mathbf{x}^{(i)}$ ($i=2,3,4$). Thus we have $\mathrm{Div}
\mathbf{S} = O(Z^3)$, which implies that the first term is of
$O(h^4)$. While $(\ref{eqX1})$, which is used to eliminate
$\mathbf{x}^{(1)}$, makes the second term be exactly zero. The error
of the third term is $O(h^4)$ due to $(\ref{eqX2})$. Actually, it is
easy to see the 3-D field equations and traction conditions on the top
and bottom surfaces are satisfied up to $O(h^2)$ in a pointwise
manner. Thus, we conclude that, except the two edge terms (the fourth
and fifth terms in $\delta E$), the present plate equation together
with the intrinsic relations among $\mathbf{x}^{(i)}$ ($i=0,...,4$)
guarantee an $O(h^4)$ error for for the first three terms in $\delta
E$.  Next, we shall introduce proper plate boundary conditions to make
the two edge terms be $O(h^4)$.

\remark In existing plate theories, usually the series expansion about
the middle surface is used, perhaps due to some symmetry properties
and separation of bending and stretching deformations in the
derivation. Here we abandon this convention and adopt an expansion
about the bottom surface, which enables us to derive the exact
relation between $\mathbf{x}^{(1)}$ and $\mathbf{x}^{(0)}$, leading to
a simple system for only $\mathbf{x}^{(0)}$. If we follow the
middle-surface expansion (as we have attempted initially), the
consistent 2-D plate equations with the same $O(h^4)$ error for
$\delta E$ will be a coupled system for $\mathbf{x}^{(1)}$ and
$\mathbf{x}^{(0)}$ (much more complicated). In the 2-D energy approach
to be presented in section 5, the middle-surface expansion also causes
the difficulty in finding accurate enough relation between
$\mathbf{x}^{(1)}$ and $\mathbf{x}^{(0)}$.

\subsection{Boundary conditions}

In this subsection, we aim to reduce the 3-D boundary conditions to
appropriate ones for the derived 2-D vector plate equation. Since the
plate equation is of fourth order, two conditions regarding
$\mathbf{x}^{(0)}$ or its derivatives are needed, either on the
position boundary $\partial \Omega_0$ or on the traction boundary
$\partial \Omega_q$.

\vspace{0.3cm} \noindent {\bf Case 1. Prescribed position in the 3-D
formulation}

Suppose that on $\partial \Omega_0 \times [0,2h]$ the position
$\mathbf{b}$ is prescribed. In this case, in order to satisfy the
consistency criterion, for the 2-D vector plate equation we adopt the
following two conditions
\begin{equation}
\label{eq18}
\begin{aligned}
& \mathbf{x}^{(0)} = \mathbf{b}^{(0)}(s),\quad \bar{\mathbf{x}} =
\bar{\mathbf{b}} \quad on \quad \partial \Omega_{q},\\
\Leftrightarrow \quad & \mathbf{x}^{(0)} = \mathbf{b}^{(0)}(s),\quad
\mathbf{x}^{(1)} + \frac{2}{3}h \mathbf{x}^{(2)}+ \frac{1}{3}h^2
\mathbf{x}^{(3)}+O(h^3) =
\frac{1}{h}(\bar{\mathbf{b}}-\mathbf{b}^{(0)}),
\end{aligned}
\end{equation}
where a bar over a quantity represents the through-thickness average
and $\mathbf{b}^{(0)} = \mathbf{b}|_{Z=0}$. The second condition
contains up to the third-order derivatives of $\mathbf{x}^{(0)}$ upon
using the recursion relations. We point out that the first condition can be replaced by prescribing the position at any given point.

To check the consistency,  we examine the asymptotic order of the
fourth term in $(\ref{eq2})$:
\begin{equation}
\label{eq19}
\begin{aligned}
\int_{0}^{2h} \mathbf{S}^T \mathbf{N} \cdot \delta
\mathbf{x}\mathrm{dZ}=& \int_{0}^{2h} \mathbf{S}^{(0)T} \mathbf{N}
\cdot \delta \mathbf{x}\mathrm{dZ} + \int_{0}^{2h} Z \mathbf{S}^{(1)T}
\mathbf{N} \cdot [\delta \mathbf{x}^{(0)} + Z \delta
\mathbf{x}^{(1)}]\mathrm{dZ}\\
&+\int_{0}^{2h} \frac{1}{2}Z^2 \mathbf{S}^{(2)T} \mathbf{N} \cdot
\delta \mathbf{x}^{(0)}\mathrm{dZ} + O(h^4).
\end{aligned}
\end{equation}
Obviously the first term in $(\ref{eq19})$ is zero due to the
condition $(\ref{eq18})_2$. The last term is zero as
$\delta\mathbf{x}^{(0)}=0$, and the second term is of $O(h^4)$ since
it is easy to see from $(\ref{eq18})_2$ that $\delta
\mathbf{x}^{(1)}=O(h)$.

\vspace{0.3cm} \noindent {\bf Case 2. Prescribed traction in the 3-D
formulation}

Suppose that on $\partial \Omega_{q}\times [0,2h]$ the traction
$\mathbf{q}$ is specified and is $C^4$ in $Z$. Denote the coefficients
of its four-term Taylor expansion by $\mathbf{q}^{(i)}$ ($i=0,1,2,3$).
In this case, we adopt the following two conditions
\begin{equation}
\label{eq20}
\begin{aligned}
&\frac{1}{2 h} \int_0^{2 h} \mathbf{S}^{T} \mathbf{N} \mathrm{dZ
}=\frac{1}{2 h} \int_0^{2 h}  \mathbf{q} \mathrm{dZ
} = {\mathbf{q}_0},\quad  \mathbf{S}^{(0)T}\mathbf{N}  =\mathbf{q}^{(0)}\\
\Leftrightarrow \quad  &  \bar{\mathbf{S}}^T \mathbf{N} =\left[
\mathbf{S}^{(0)} + h  \mathbf{S}^{(1)} + \frac{2}{3} h^2
\mathbf{S}^{(2)} +\frac{1}{3} h^3 \mathbf{S}^{(3)} + O(h^3)\right]^T
\mathbf{N} = {\mathbf{q}_0}, \quad \mathbf{S}^{(0)T}\mathbf{N}
=\mathbf{q}^{(0)},
\end{aligned}
\end{equation}
where ${\mathbf{q}_0}$ is the averaged traction and can be expressed
in terms of $\mathbf{q}^{(i)}$ in the same way as the left-hand side.
Similarly the second traction condition at $Z=0$ may be replaced by
one at an arbitrary $Z$. Alternatively, the second condition can be
suitably replaced by the specified moment about the middle line as
follows
\begin{equation}
\label{eq21}
\begin{aligned}
&\frac{1}{2 h} \int_0^{2 h} (Z-h) \mathbf{S}^{T} \mathbf{N} \mathrm{dZ
}=\frac{1}{2 h} \int_0^{2h} (Z-h)\mathbf{q} \mathrm{dZ}=\mathbf{m}_0(s),\\
\Leftrightarrow \quad  & \frac{1}{3}  \mathbf{S}^{(1)T} \mathbf{N}
+ \frac{1}{3} h \mathbf{S}^{(2)T} \mathbf{N}+\frac{1}{5} h^2
\mathbf{S}^{(3)T} \mathbf{N}+O(h^3) =h^-2 \mathbf{m}_0(s),
\end{aligned}
\end{equation}
where $\mathbf{m}_0(s)$ can be expressed in terms of
$\mathbf{q}^{(i)}$ in the same way as the
left-hand side. The first two components of $2h \mathbf{m}_0$ are the
classical bending moment and the twisting moment respectively
\cite{kienzler2002}. The third component does not has a clear physical
meaning (somehow related to the extension of the edge cross-section
along $Z$ direction), which can also replaced by the third component
of $(\ref{eq20})_2$ whose physical meaning is clear. Also,
$\mathbf{S}^{(3)T} \mathbf{N}$ terms are kept in order to make the
boundary conditions up to $O(h^2)$ correct.

To check the consistency, we examine the asymptotic order of the fifth
term in $(\ref{eq2})$. For convenience, we denote $\tilde{\mathbf{q}}
=\mathbf{S}^T  \mathbf{N} -\mathbf{q}$ and use
$\mathbf{\tilde{q}}^{(i)}$ ($i=0,1,2$) to represent the coefficients
of its Taylor expansion . We have
\begin{equation}
\label{eq22} \int_{0}^{2h} \tilde{\mathbf{q}} \cdot \delta \mathbf{x}
\mathrm{dZ} =\int_{0}^{2h} \tilde{\mathbf{q}} \cdot \delta
\mathbf{x}^{(0)} \mathrm{dZ} + \int_{0}^{2h} Z
\left[\tilde{\mathbf{q}}^{(0)} + Z \tilde{\mathbf{q}}^{(1)}\right]
\cdot \delta \mathbf{x}^{(1)}\mathrm{dZ}+\int_{0}^{2h} \frac{1}{2}Z^2
\tilde{\mathbf{q}}^{(0)} \cdot \delta \mathbf{x}^{(2)}\mathrm{dZ} +
O(h^4).
\end{equation}
The first term is zero due to the condition $(\ref{eq20})_1$. By
simple manipulations of the two conditions in $(\ref{eq20})$ (or
$(\ref{eq20})_1$ and $(\ref{eq21})$), one can show
$\tilde{\mathbf{q}}^{(0)}=O(h^2)$ and $\tilde{\mathbf{q}}^{(1)}=O(h)$,
and consequently the remaining terms are at least $O(h^4)$.

\remark  Depending on the problems, some combinations of
$(\ref{eq18})_{1,2}$ and $(\ref{eq20})_{1,2}$ or $(\ref{eq21})$ can be
used. It should also be pointed that the above proposed boundary
conditions do not satisfy edge boundary conditions $(\ref{eq3})_{4,5}$
in the 3-D differential formulation to the required order in a
pointwise manner about $Z$. Rather, they are satisfied in certain
average manners. As a result, locally near the edge the plate solution
may not achieve $O(h^2)$ accuracy, especially when a boundary layer is
present. We should mention that most common plate boundary conditions
are derived from the variational (or virtual work) principle with the
introduction of generalized traction and bending moment, which do not
conform with those in the 3-D formulation. Here, when the 3-D position
or traction conditions are known, we do not need such artificial
quantities for plate boundary conditions. However, when the 3-D
position or traction conditions are not known, as a price to pay for
such an uncertainty, they will be needed, as will be seen later.

To sum up, the 2-D vector plate equation $(\ref{eq17})_1$ together
with boundary conditions $(\ref{eq18})$ and $(\ref{eq20})$ (or
$(\ref{eq20})_1$ and $(\ref{eq21})$), ensure each term in the
variation $\delta E$ to be either zero or $O(h^4)$. To our knowledge,
no existing plate models enjoy such a consistency. Also, the derived
plate theory does not need to introduce either artificial quantities
like generalized traction and bending moment nor unphysical
higher-order (generalized) stress resultants, which are often present
in existing plate theories.

\section{Associated weak formulations}

In this section, we deduce the associated weak formulations for the
previous 2-D plate system in a way similar to that in
\cite{Ogden1984}, in order to derive formulations suitable for
numerical calculations. Another purpose is to introduce suitable
boundary conditions for a number of practical cases that the 3-D edge
conditions are not known (e.g., for a pinned edge one does not know
the traction distribution).

First, multiplying both sides of the 2-D plate equation
$(\ref{eq17})_1$ by $\xi= \delta \mathbf{x}^{(0)}$, we obtain
\begin{equation}
\label{eq23} \int_{\Omega}  (\nabla \cdot \bar{\mathbf{S}})\cdot \xi
\mathrm{d\mathbf{r}} = - \int_{\Omega}  \bar{\mathbf{q}} \cdot \xi
\mathrm{d\mathbf{r}}\quad \Rightarrow \quad \int_{\partial \Omega}
(\bar{\mathbf{S}}^T \mathbf{N})\cdot \xi \mathrm{ds} - \int_{\Omega}
\bar{\mathbf{S}}: \nabla \xi \mathrm{d\mathbf{r}} = - \int_{\Omega}
\bar{\mathbf{q}} \cdot \xi \mathrm{d\mathbf{r}}.
\end{equation}
In general, for a fourth-order differential system, the weak
formulation should only contain up to the second-order derivative of
$\mathbf{x}^{(0)}$ (especially regarding the functional space in
finite element calculations). However, $\bar{\mathbf{S}}$ involves the
third-order derivative of $\mathbf{x}^{(0)}$, which we intend to
eliminate. This term originates from the term $\mathbf{F}^{(2)}$ in
$\mathbf{S}^{(2)}$ (see $(\ref{eq10})_3$ and $(\ref{eq8})$), which is
decomposed into two parts for identification
\begin{equation}
\label{eq24}
\begin{aligned}
\mathbf{F}^{(2)} &= \mathbf{F}_1^{(2)} + \mathbf{F}_2^{(2)},\quad
\mathbf{F}_1^{(2)} =  -\left\{\mathbf{B}^{-1}({\mathcal{A}}^2
[\mathbf{F}^{(1)},\mathbf{F}^{(1)}])^T \mathbf{k} \right\}\otimes \mathbf{k},\\
\mathbf{F}_2^{(2)} &= \nabla \mathbf{x}^{(2)} -\left\{\mathbf{B}^{-1}
\left(({\mathcal{A}}^1 [\nabla \mathbf{x}^{(2)}])^T \mathbf{k} +
\nabla \cdot \mathbf{S}^{(1)}\right) \right\}\otimes \mathbf{k},
\end{aligned}
\end{equation}
where only the second part needs special attention. Correspondingly,
we have
\begin{equation}
\label{eq25}
\begin{aligned}
& \bar{\mathbf{S}} : \nabla \xi  = \mathcal{W}_1 (\nabla \nabla
\mathbf{x} ^{(0)}, \nabla \xi) + \mathcal{W}_2 (\nabla \nabla \nabla
\mathbf{x} ^{(0)}, \nabla \xi),
\end{aligned}
\end{equation}
where only the highest-order derivative is listed in the arguments.
Curlicue letters including $\mathcal{W}_i$ ($i=1,2,3$) and the
following $\mathbf{\mathcal{S}}_0,\eta$ are used to denote quantities
involving the virtual variable $\xi$. Explicitly, we have
\begin{equation}
\label{eq26}
\begin{aligned}
\mathcal{W}_1  & =  (\mathbf{S}^{(0)} + h  \mathbf{S}^{(1)}): \nabla
\xi + \frac{2}{3} h^2 {\mathcal{A}}^2 [\mathbf{F}^{(1)},
\mathbf{F}^{(1)}] : \nabla \xi + \frac{2}{3} h^2 {\mathcal{A}}^1
[\mathbf{F}_1^{(2)}] : \nabla \xi,\\
\mathcal{W}_2  &=\frac{2}{3} h^2  {\mathcal{A}}^1 [\mathbf{F}_2^{(2)}]
: \nabla \xi = \frac{2}{3} h^2  {\mathcal{A}}^1 [\nabla \xi] :
\mathbf{F}_2^{(2)}=\frac{2}{3} h^2 \left(\mathbf{\mathcal{S}}_0 :
\nabla \mathbf{x}^{(2)} + \eta \cdot ( \nabla \cdot
\mathbf{S}^{(1)})\right),
\end{aligned}
\end{equation}
where the symmetry property of ${\mathcal{A}}^1$ has been utilized in
deriving $(\ref{eq26})_2$, and the quantities $\mathbf{\mathcal{S}}_0$
and $\eta$ are defined as
\begin{equation}
\label{eq27}
\begin{aligned}
\eta= - \mathbf{B}^{-1} ({\mathcal{A}}^1 [\nabla \xi])^T \mathbf{k}
,\quad \mathbf{\mathcal{S}}_0 = {\mathcal{A}}^1 [\nabla \xi + \eta
\otimes \mathbf{k}].
\end{aligned}
\end{equation}
These quantities are actually the variations of $\mathbf{x}^{(1)}$ and
$\mathbf{S}^{(0)}$ respectively, since from $(\ref{eq10})_1$ and
$(\ref{eqX1})$ one can deduce
\begin{equation}
\label{eq28}
\begin{aligned}
\eta=\delta \mathbf{x}^{(1)},\quad \mathbf{\mathcal{S}}_0 = \delta
\mathbf{S}^{(0)}.
\end{aligned}
\end{equation}
Then, the third-order derivative terms in $(\ref{eq23})$ is eliminated
by the divergence theorem
\begin{equation}
\label{eq29}
\begin{aligned}
&\int_\Omega \mathcal{W}_2 \mathrm{d\mathbf{r}} = \frac{2}{3} h^2
\int_{\partial \Omega} (\mathbf{\mathcal{S}}_0^{T} \mathbf{N})\cdot
\mathbf{x}^{(2)} + (\mathbf{S}^{(1)T} \mathbf{N})\cdot \eta
\mathrm{ds} + \int_{\Omega} \mathcal{W}_3 \mathrm{d\mathbf{r}}, \\
&\mathcal{W}_3 (\nabla \nabla \mathbf{x}^{(0)}, \nabla \nabla \xi)=
-\frac{2}{3} h^2 \left( (\nabla \cdot { \mathbf{\mathcal{S}}}_0)\cdot
\mathbf{x}^{(2)} + \mathbf{S}^{(1)} : \nabla \eta\right).
\end{aligned}
\end{equation}
In summary, we have the following 2-D weak form
\begin{equation}
\label{eq30}
\begin{aligned}
& \int_{\Omega} \left(\mathcal{W}_1 + \mathcal{W}_3 -\bar{\mathbf{q}}
\cdot \xi \right)\mathrm{d\mathbf{r}} = \int_{\partial \Omega} \left(
\bar{\mathbf{S}}^T \mathbf{N}\cdot \xi - \frac{2}{3} h^2
\mathbf{S}^{(1)T} \mathbf{N}\cdot \eta -\frac{2}{3} h^2
\mathbf{\mathcal{S}}_0^{T} \mathbf{N} \cdot \mathbf{x}^{(2)} \right)
\mathrm{ds}.
\end{aligned}
\end{equation}
In the following, we will rewrite $(\ref{eq30})$ for two distinct
situations, according to different types of edge boundary conditions.

\vspace{0.2cm} \noindent {\bf (1) Edge position and traction in the
3-D formulation are known}

In the differential formulation of the plate system, the suitable
boundary conditions for this situation have been introduced in  case 1
and case 2 of subsection 3(c) (respectively on $\partial \Omega_0$ and
$\partial \Omega_q$). Now we use them to further simplify the the
above weak formulation.

On $\partial \Omega_0$,  $(\ref{eq18})$ holds, and from which it is
easy to deduce $\xi=\delta \mathbf{x^{(0)}}=0, \eta = \delta
\mathbf{x}^{(1)} =O(h)$. The latter  and $(\ref{eq27})_1$ imply that
$\nabla \xi = O(h)$, which, together with $(\ref{eq27})_2$, further
imply $\mathbf{\mathcal{S}}_0= O(h)$. As a result, the part on
$\partial \Omega_0$ of the right-hand side of $(\ref{eq30})$ is of
$O(h^3)$ and is thus ignored (note that a factor $2h$ is divided in
deriving the plate equation). In this context, we can readily replace
$\partial \Omega$ by $\partial \Omega_q$ for the edge boundary
integral.

While on $\partial \Omega_q$, from the boundary conditions in case 2
it is easy to deduce $\mathbf{ \mathcal{S} }_0^{T} \mathbf{N} = \delta
[\mathbf{S}^{(0)T} \mathbf{N}] =O(h^2)$. Thus, the third term inside
the boundary integral can be neglected. Then, the 2-D weak form
$(\ref{eq30})$ reduces to
\begin{equation}
\label{eq31}
\begin{aligned}
\int_{\Omega} \left(\mathcal{W}_1 + \mathcal{W}_3 -\bar{\mathbf{q}}
\cdot \xi \right)\mathrm{d\mathbf{r}} = \int_{\partial \Omega_q}
\left( \bar{\mathbf{S}}^T \mathbf{N}\cdot \xi - \frac{2}{3} h^2
\mathbf{S}^{(1)T} \mathbf{N}\cdot \eta \right) \mathrm{ds}=
\int_{\partial \Omega_q}  {\mathbf{q}_0} \cdot \xi - 2 \mathbf{m}_0
\cdot \eta  \mathrm{ds}.
\end{aligned}
\end{equation}
in which the second equality is obtained upon using $(\ref{eq20})_1$
and $(\ref{eq21})$ to replace the two traction terms and neglecting
$O(h^3)$ terms. The above weak form is suitable for finite element
calculations with prescribed ${\mathbf{q}_0}$ and $\mathbf{m}_0$
(which are known once the edge traction in the 3-D formulation is
given). It should be noted that attention should be paid on the
suitable functional space for the test functions, which should conform
with the restrictions $(\ref{eq18})$ on $\partial \Omega_0$.

\vspace{0.2cm} \noindent { \bf (2) Edge position and traction in the
3-D formulation are unknown}

In a number of practical situations, one does not know the edge
traction distribution (e.g. a pinned edge) or displacement
distribution (e.g. a clamped edge). In these cases, it does not make
sense to introduce plate boundary conditions for the purpose of making
the two edge terms in $\delta E$  (see $(\ref{eq2})$) be of certain
consistent order. Rather, one can propose the so-called natural
boundary conditions according to the weak formulation, and most
existing plate models deduce boundary conditions in this way. To this
end, we would like to rewrite the boundary integral in (4.8) by using
$\xi_{,N}$, the normal derivative of $\xi$.

For convenience,  we introduce a third-order (moment) tensor
$\mathbf{M}= \mathbf{M}(\nabla \mathbf{x}^{(0)} ,\nabla \nabla
\mathbf{x}^{(0)})$ according to the last two terms in the boundary
integral of (4.8):
\begin{equation}
\label{eq32}
\begin{aligned}
& -\frac{2}{3} h^2 \left( \mathbf{S}^{(1)T} \mathbf{N}\cdot \eta +
\mathbf{\mathcal{S}}_0^{T}
\mathbf{N} \cdot \mathbf{x}^{(2)}\right)=(\mathbf{M [N]})^T: \nabla \xi.\\
\end{aligned}
\end{equation}
And further we introduce the decomposition
\begin{equation}
\label{eq32_1}
\begin{aligned}
\nabla \xi =  \xi_{,s} \otimes \mathbf{T} + \xi_{,N} \otimes \mathbf{N},\\
\end{aligned}
\end{equation}
where $\mathbf{T}$ and $\xi_{,s}$ are respectively the unit tangential
vector and tangent derivative. Substituting these two relations
$(\ref{eq32},\ref{eq32_1})$ into the boundary integral in
$(\ref{eq30})$ and a simple integration by part leads to (possible
corner forces are not considered here)
\begin{equation}
\label{eq33}
\begin{aligned}
& \int_{\Omega} \left(\mathcal{W}_1 + \mathcal{W}_3 -\bar{\mathbf{q}}
\cdot \xi \right)\mathrm{d\mathbf{r}} = \int_{\partial \Omega}
\left(\bar{\mathbf{S}}^T \mathbf{N}- (\mathbf{M [N,T]})_{,s}
\right)\cdot \xi + \mathbf{M [N,N]}\cdot
\mathbf{\xi}_{,N}  \mathrm{ds},\\
\end{aligned}
\end{equation}
where the two quantities relating to $\mathbf{M}$ are given by
\begin{equation}
\label{eq34}
\begin{aligned}
& \mathbf{M [N,N]} = \frac{2}{3} h^2  \left(\mathbf{B}_1^T
\mathbf{B}^{-1} \left(\mathbf{S}^{(1)T} \mathbf{N} + \mathbf{B}_1
\mathbf{x}^{(2)}\right) - \mathbf{B}_2 \mathbf{x}^{(2)}\right),\\
& \mathbf{M [N,T]} =  \frac{2}{3} h^2  \left( \mathbf{B}_3^T
\mathbf{B}^{-1} \left(\mathbf{S}^{(1)T} \mathbf{N} + \mathbf{B}_1
\mathbf{x}^{(2)}\right) - \mathbf{B}_4 \mathbf{x}^{(2)}\right),\\
&(B_1)_{ij}= {\mathcal{A}}^1_{3i\alpha j} N_\alpha,\quad ({B}_2)_{ij}=
{\mathcal{A}}^1_{\alpha i\beta j} N_\alpha N_\beta,({B}_3)_{ij}=
{\mathcal{A}}^1_{3i\alpha j} T_\alpha,\quad ({B}_4)_{ij}=
{\mathcal{A}}^1_{\alpha i\beta j} T_\alpha N_\beta.
\end{aligned}
\end{equation}
In the standard plate theory (see \cite{reddy2011,steigmann2014}), the
terms before $\xi$ and $\xi_{,N}$ are considered as the generalized
average traction and generalized bending moment respectively.

Formally, if we regard $\mathcal{W} =:\mathcal{W}_1+ \mathcal{W}_3$ as
the variation (increase) of the plate stress work due to the virtual
position/displacement $\xi$, the weak form $(\ref{eq33})$ can be
rewritten as
\begin{equation}
\label{eq36}
\begin{aligned}
& \int_{\Omega} \mathcal{W} \mathrm{d\mathbf{r}} = \int_{\Omega}
\bar{\mathbf{q}} \cdot \xi \mathrm{d\mathbf{r}} + \int_{\partial
\Omega} \hat{\mathbf{q}}(s) \cdot \xi + \hat{\mathbf{m}}_0(s) \cdot
\mathbf{\xi}_{,N} \mathrm{ds},
\end{aligned}
\end{equation}
where $\hat{\mathbf{q}}$ and $\hat{\mathbf{m}_0}$ are respectively the
applied generalized traction and bending moment at the edge. Since the
three terms on the right-hand side are respectively the virtual work
by the plate body force (caused by the tractions of top and bottom
surfaces), by the generalized edge traction and the generalized edge
bending moment, the above equation is simply the 2-D virtual work
principle for the plate. This weak form can be used for implementing
finite element schemes.

In the expressions of $\mathcal{W}_1 $ and $\mathcal{W}_3$ (see
$(\ref{eq26})_1$ and $(\ref{eq29})_2$), except that the
$\mathbf{S}^{(0)}$-term depends on $\nabla \mathbf{x}^{(0)}$ only, the
remaining terms depend on both $\nabla \mathbf{x}^{(0)}$ and $\nabla
\nabla \mathbf{x}^{(0)}$. So, it is difficult to identify which part
representing stretching and which part representing bending, rather
these two effects are combined together in the variation of the stress
work. Also, if there exists an function $W(\nabla
\mathbf{x}^{(0)},\nabla \nabla \mathbf{x}^{(0)})$ such that
$\int_{\Omega} \delta W \mathrm{d\mathbf{r}} = \int_{\Omega}
\mathcal{W} \mathrm{d\mathbf{r}}$,  then $W$ can be regarded as the
2-D plate strain energy function. But, in general, although it is not
a proof, we expect that such a function does not exist. One reason is
that $W$ is an energy not only depending on the material and plate
thickness but also on the applied traction $\mathbf{q}^{-}$ on the
bottom surface ($\mathbf{x^{(1)}}$ is solved in terms of $\nabla
\mathbf{x}^{(0)}$ and $\mathbf{q}^{-}$, see $(\ref{eqX1})$). From this
point view, the derived plate system may not be treated as an energy
minimization problem.

Based on the weak form $(\ref{eq33})$, we are ready to introduce the
suitable boundary conditions for various practical cases.

\vspace{0.3cm} \noindent {\bf Case 3. A clamped edge}

For a clamped edge, the bottom position and its normal derivative are
assigned, that is
\begin{equation}
\label{eq38}
\begin{aligned}
\mathbf{x}^{(0)}= \mathbf{b}^{(0)}(s),\quad
\mathbf{x}_{,N}^{(0)}=\hat{\mathbf{b}}^{(1)}(s).
\end{aligned}
\end{equation}
These two conditions can be recast in terms of the displacement
according to the relation of $\mathbf{u}^{(0)} = \mathbf{x}^{(0)} -
[\mathbf{r},0]^T$, since sometimes it is convenient to use
displacement directly, e.g. $\mathbf{u}_{,N}^{(0)}=\mathbf{0}$.

\vspace{0.3cm} \noindent {\bf Case 4. Prescribed generalized traction
and bending moment}

In this case, the boundary conditions are
\begin{equation}
\label{eq40}
\begin{aligned}
&\bar{\mathbf{S}}^T \mathbf{N}- (\mathbf{M [N,T]})_{,s} =
\hat{\mathbf{q}}(s),\quad \mathbf{M [N,N]}=\hat{\mathbf{m}}_0(s).
\end{aligned}
\end{equation}

\vspace{0.3cm} \noindent {\bf Case 5. A pinned edge}

In this case, the position is assigned and the generalized bending
moment vanishes. Correspondingly, we have
\begin{equation}
\label{eq41}
\begin{aligned}
\mathbf{x}^{(0)} (s) = \mathbf{b}^{(0)}(s),\quad \mathbf{M
[N,N]}=\mathbf{0}.
\end{aligned}
\end{equation}

\vspace{0.3cm} \noindent {\bf Case 6. A simply-supported edge}

We take the left edge $X_1=0$ of a rectangular plate with
$\mathbf{X}=(X_1,X_2,Z)$ for example. The normal vector is
$\mathbf{N}=(-1,0,0)$ and the variable along the edge is $s=-X_2$.
Then, the boundary conditions are
\begin{equation}
\label{eq42}
\begin{aligned}
&\mathbf{u}_2^{(0)} (0,X_2) = \mathbf{u}_3^{(0)} (0,X_2) =0,\quad
(\bar{\mathbf{S}}^T \mathbf{N})_1+  (\mathbf{M
[N,T]})_{1,2} = 0,\\
& (\mathbf{M [N,N]})_2=(\mathbf{M [N,N]})_3=0,\quad
\mathbf{u}_{1,1}^{(0)} (0,X_2) =0.
\end{aligned}
\end{equation}
For a plate with other geometries, we should properly modify
$(\ref{eq42})$ according to its tangential and normal directions.

\section{A 2-D plate strain energy function}

The previous derived plate system has the force-balance structure in
all three directions and is consistent with the principle of
stationary 3-D potential energy up to $O(h^3)$, but a price to pay is
that it may not be treated as an energy minimization problem. In this
section, we shall somewhat relax the consistency requirement to obtain
a 2-D strain energy function for the plate in a way similar to that
used in \cite{steigmann2013}. In that paper, the case of zero (or
small enough) tractions on the top and bottom surfaces  was considered
and, as pointed out by the author, a non-negligible error was incurred
when $\mathbf{x^{(1)}}$ was expressed in terms of $\mathbf{x^{(0)}}$.
While here we do not put any restrictions on the applied tractions and
the derivation avoids such an error. We shall first truncate the 3-D
potential energy $(\ref{eq1})$ to $O(h^3)$ and then simplify it to a
2-D potential energy by eliminating the third-order derivatives. The
details are described below.

Similar to $(\ref{eq4})$, the strain energy is also expanded as
\begin{equation}
\label{eq43}
\begin{aligned}
&\Phi (\mathbf{F}) = \Phi^{(0)}(\mathbf{r}) + Z \Phi^{(1)}(\mathbf{r})
+ \frac{1}{2} Z^2 \Phi^{(2)} (\mathbf{r}) +\cdots, \quad 0\le Z\le 2h.
\end{aligned}
\end{equation}
By the chain rule with the help of $(\ref{eq7}, \ref{eq9})$, one can
deduce the relations
\begin{equation}
\label{eq43_1}
\begin{aligned}
\Phi^{(0)} =\Phi (\mathbf{F}^{(0)}), \quad \Phi^{(1)}
=\mathbf{S}^{(0)} : \mathbf{F}^{(1)}, \quad \Phi^{(2)} =
\mathbf{S}^{(0)} : \mathbf{F}^{(2)} +  \mathbf{S}^{(1)}:
\mathbf{F}^{(1)}.
\end{aligned}
\end{equation}
Subsequently with the series expansions $(\ref{eq4},\ref{eq43})$ and
by an integration, the two parts in the 3-D potential energy $E$ given
in $(\ref{eq1})$ are truncated as
\begin{equation}
\label{eq44}
\begin{aligned}
\bar{\Phi}  =&  2 h \int_{\Omega}  \left( \Phi^{(0)}(\mathbf{r}) + h
\Phi^{(1)}(\mathbf{r}) + \frac{2 }{3} h^2 \Phi^{(2)}
(\mathbf{r})\right) \mathrm{d\mathbf{r}} + O(h^4)\\
\bar{V} = &  2h  \int_{\Omega} \bar{\mathbf{q}} \cdot \mathbf{x}^{(0)}
+ \mathbf{q}^{+} \cdot \left(\mathbf{x}^{(1)} + h \mathbf{x}^{(2)} +
\frac{2}{3} h^2 \mathbf{x}^{(3)}\right)
\mathrm{d\mathbf{r}}\\
& + 2h \int_{\partial \Omega_q} \left( \mathbf{q}_0 \cdot
\mathbf{x}^{(0)} + h \mathbf{q}_1 \cdot \mathbf{x}^{(1)}  +
\frac{2}{3} h^2 \mathbf{q}^{(0)} \cdot \mathbf{x}^{(2)}\right)
\mathrm{ds} + O(h^4),
\end{aligned}
\end{equation}
where $\mathbf{\bar{q}}$, $\mathbf{q}_0$ and $\mathbf{q}^{(0)}$ have
the same meanings defined previously, and $\mathbf{q}_1$ is defined as
$\frac{1}{2 h^2} \int_0^{2h} Z \mathbf{q} dZ$.

The above terms in $(\ref{eq44})$ involve quantities from
$\mathbf{x}^{(0)}$ to $\mathbf{x}^{(3)}$, which are not independent in
the 2-D formulation since $\mathbf{x}^{(i)} (i=1,2,3)$ can be
expressed in terms of $\mathbf{x}^{(0)}$ according to the 3-D
formulation as done in section 3. Therefore, those relations should be
used before taking the variation. As a result, the strain energy
$\bar{\Phi}$ will involve up to the third-order derivatives of
$\mathbf{x} ^{(0)}$. To eliminate them, similar to that in section 4,
we further simplify the following term in $\Phi^{(2)}$ by the
divergence theorem
\begin{equation}
\label{eq46}
\begin{aligned}
&\int_\Omega \mathbf{S}^{(0)} : \mathbf{F}^{(2)} \mathrm{d\mathbf{r}}
= \int_\Omega \mathbf{S}^{(0)} : \nabla \mathbf{x}^{(2)} +
(\mathbf{S}^{(0)T} \mathbf{k}) \cdot \mathbf{x}^{(3)}
\mathrm{d\mathbf{r}}\\
=& \int_\Omega - (\nabla \cdot \mathbf{S}^{(0)}) \cdot
\mathbf{x}^{(2)} + (\mathbf{S}^{(0)T} \mathbf{k}) \cdot
\mathbf{x}^{(3)} \mathrm{d\mathbf{r}} + \int_{\partial \Omega}
(\mathbf{S}^{(0)T} \mathbf{N}) \cdot \mathbf{x}^{(2)} ds
\end{aligned}
\end{equation}
Combining the last two terms in $(\ref{eq46})$ and the two $O(h^3)$
terms of $\bar{V}$ in $(\ref{eq44})_2$ (upon using $(\ref{eqX1})$)
leads to
\begin{equation}
\label{eq47}
\begin{aligned}
& \frac{8}{3} h^4  \int_\Omega \bar{\mathbf{q}} \cdot \mathbf{x}^{(3)}
\mathrm{d\mathbf{r}} + \frac{4}{3} h^3 \int_{\partial \Omega_q}
(\mathbf{S}^{(0)T} \mathbf{N} -\mathbf{q}^{(0)}) \cdot
\mathbf{x}^{(2)} ds + \frac{4}{3} h^3 \int_{\partial \Omega_0}
\mathbf{S}^{(0)T} \mathbf{N} \cdot \mathbf{x}^{(2)} ds.
\end{aligned}
\end{equation}
The first term can be neglected as it is $O(h^4)$. The second term is
on the traction boundary $\partial\Omega_q$, and for any reasonably
prescribed traction conditions the term $(\mathbf{S}^{(0)T} \mathbf{N}
-\mathbf{q}^{(0)})$ should be at least $O(h)$ (cf. $(\ref{eq51})$ in
Appendix A). Thus, this term can also be neglected.  We shall also
neglect the third term, which represents the contribution from
$\mathbf{x}^{(2)}$ on $\partial\Omega_0$. In terms of asymptotic
order, one cannot say such an approximation is consistent as it is
$O(h^3)$. Nevertheless, we drop it based on the following reasons.
Firstly, the 3-D edge conditions on $\Omega_0$ are not known in some
practical problems (see section 4), and in this case the error to any
proposed plate boundary conditions can never be estimated. So, the
above approximation does not worsen the situation. Secondly, no
suitable boundary conditions can be prescribed if this term is kept,
since three boundary conditions on $\partial \Omega_0$ (inconsistent
with the required two) will be needed after the 2-D variational
principle is performed. Thus, this term leads to an essential and
inevitable difficulty on the boundary conditions in a 2-D energy
formulation. This point was already noticed by Steigmann
\cite{steigmann2008}, who ignored such terms in his plate theory by
stipulating that the {\it posteriori} boundary data of
$\mathbf{x}^{(2)}$ is consistently prescribed.

With the above-mentioned approximation, the  2-D potential energy
reduces to
\begin{equation}
\label{eq48}
\begin{aligned}
E_{2D}= &\bar{\Phi}_{2D} - \bar{V}_{2D} = \int_\Omega \Phi_{2D}
\mathrm{d\mathbf{r}} - 2h \int_\Omega \bar{\mathbf{q}} \cdot
\mathbf{x}^{(0)} \mathrm{d\mathbf{r}} - 2h \int_{\partial \Omega_q}
\left( \mathbf{q}_0 \cdot \mathbf{x}^{(0)} + h \mathbf{q}_1
\cdot \mathbf{x}^{(1)}\right)ds,\\
\Phi_{2D} =& 2 h  \Phi^{(0)} +2 h^2 \Phi^{(1)} + \frac{4 }{3} h^3
\left(\mathbf{S}^{(1)}: \mathbf{F}^{(1)} - (\nabla \cdot
\mathbf{S}^{(0)}) \cdot \mathbf{x}^{(2)}\right) - 2h \mathbf{q}^{+}
\cdot \left(\mathbf{x}^{(1)} + h \mathbf{x}^{(2)} \right).
\end{aligned}
\end{equation}
Based on this expression, one can regard $\Phi_{2D}$ as the 2-D plate
strain energy function. We point out that it is a function not only
depending on the material and the plate thickness but also on the
applied tractions $\mathbf{q}^{-}$ and $\mathbf{q}^{+}$ (despite the
fact that the contribution from them is also present in the load
potential $\bar{V}_{2D}$). We note that $\Phi_{2D}$ is a function of
$(\nabla \mathbf{x}^{(0)}, \nabla \nabla \mathbf{x}^{(0)})$.

By taking a variation of $\bar{\Phi}_{2D}$, one can get the vector
plate equation, which is given in Appendix A. Similar to the treatment
in section 4,  by introducing the generalized average edge traction
and bending moment, it can be shown that the variation of the load
potential $\bar{V}_{2D}$ can also be written as (cf. $(\ref{eq33})$ or
$(\ref{eq36})$)
\begin{equation}
\label{eq49}
\begin{aligned}
\delta \bar{V}_{2D} = 2h\int_\Omega \bar{\mathbf{q}} \cdot \delta
\mathbf{x}^{(0)} \mathrm{d\mathbf{r}}+ 2h \int_{\partial \Omega}
\left( \hat{\mathbf{q}}_0(s) \cdot \delta \mathbf{x}^{(0)} +
\hat{\mathbf{m}}_0(s) \cdot \delta \mathbf{x}_{,N}^{(0)}\right)
\mathrm{ds}.
\end{aligned}
\end{equation}
From this form, one can easily deduce the natural boundary conditions
for the corresponding vector plate equation, which are also given in
Appendix A.

To end this section, we remark further on the
consistency/inconsistency of this 2-D energy approach for the plate
system. The second term in $\delta E$ given in $(\ref{eq2})$ is zero
as the approach still uses $(\ref{eqX1})$. Due to the dropping of the
third term in $(\ref{eq47})$, it is not possible to make the fourth
term in $\delta E$ be $O(h^4)$, and the fifth term cannot be made
$O(h^4)$ neither.  Also, we cannot show that the first and third terms
in $\delta E$ are of $O(h^4)$ separately as $\mathbf{S}^{(3)}$ is
never used in deriving $E_{2D}$  (according to the results in section
3 this coefficient is needed for each of them being this order).
However, if one considers all the terms together except the third term
in $(\ref{eq47})$ (denoted by $h^3 E^e_0$), the derivation given in
this section actually implies that $E - h^3 E^e_0=O(h^4)$, which, in
turn, implies $\delta(E - h^3 E^e_0)=O(h^4)$.  Thus, although the
present approach cannot make the first and third to fifth terms in
$\delta E$ be $O(h^4)$ separately, it can make $\delta E$ as a whole
be $O(h^4)$ when $h^3 E^e_0$ is dropped (in a traction boundary
problem $\delta \Omega_0$ is not present, and nor is this term). This
gives some justification for this 2-D energy approach, which has the
important advantage that the plate problem can be treated as an energy
minimization problem.

\section{An example: pure finite-bending problem}

To demonstrate the validity of the previously derived 2-D vector plate
equation, in this section we consider the pure bending of a
rectangular block for Hill's class of compressible materials, for
which the exact solutions are available in literature
\cite{bruhns2002,xiao2011}. Since this example is essentially a 2-D
problem and a semi-inverse one, the original system is a second-order
differential system for a scalar function. For the derived plate
theory, the final equation becomes an algebraic equation for one
scalar. In a special case, we compare the plate solution with the
exact closed-form solution.

\begin{figure}
\begin{center}
\subfigure[]{\includegraphics[width=2in]{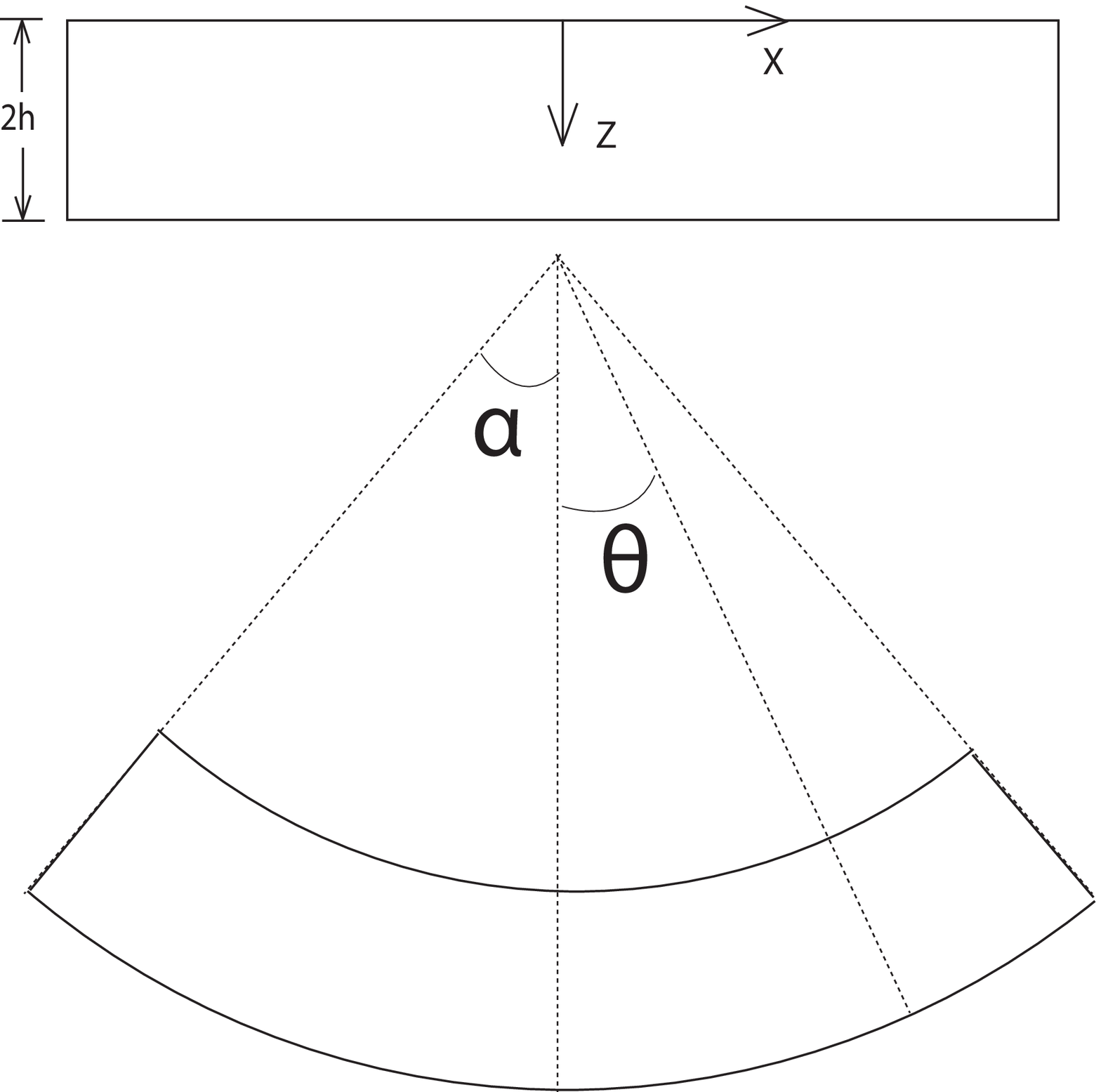}
\label{fig1a}}\hspace{1cm}
\subfigure[]{\includegraphics[width=2in]{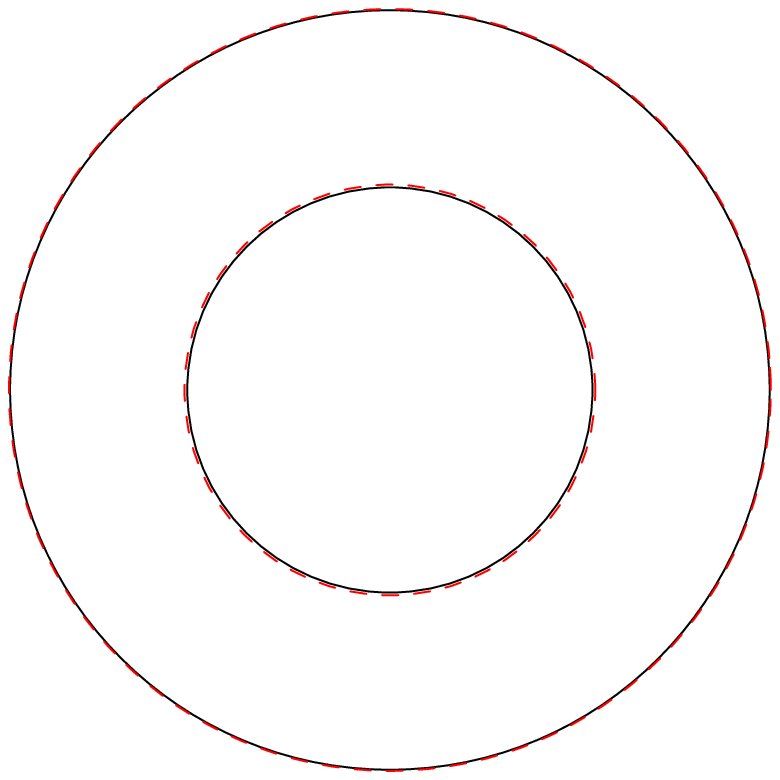} \label{fig1b}}
\end{center}
\caption{(a) The 2-D view of the reference and current states of the
block. (b) Comparison between the plate solution (dashed lines) and
exact solution (solid lines) with parameters
$\alpha=\pi,\nu=0.3,h=0.1$.} \label{fig1}
\end{figure}

Figure \ref{fig1a} schematically illustrates the 2-D picture of
reference and current states of the rectangular block. Cartesian
coordinate system $(X,Y,Z)$ is used for a reference point, while
cylindrical coordinate system $(\theta,y,r)$ is used for a current
point. Then, the bending of the block is described by
\begin{equation}
\label{eq54}
\begin{aligned}
& \theta = \frac{\alpha}{L_1} X, \qquad \quad r=r(Z), \qquad \quad y= \lambda_2 Y,\\
& -L_1 \leq X \leq L_1, \quad 0\leq Z\leq 2 h,\quad -L_2 \leq Y \leq
L_2,
\end{aligned}
\end{equation}
where $\alpha$ is the bending angle, $\lambda_2$ is the stretch normal
to the bending plane, and $L_i$ ($i=1,2$) are the two planar
dimensions. Without loss of generality we set $\lambda_2=1$ and
$L_1=1$ ($h$ is then the thickness ratio). The deformation gradient is
diagonal in the above coordinate systems
\begin{equation}
\label{eq55}
\begin{aligned}
& \mathbf{F}= \mathrm{diag}[\alpha r,\quad 1,\quad r'],
\end{aligned}
\end{equation}
where the prime denotes differentiation with respect to $Z$. As for
the constitutive relations, we recall the strain energy for Hill's
class of compressible materials \cite{hill1978},
\begin{equation}
\label{eq56}
\begin{aligned}
& \Phi = \frac{1}{2} \lambda (\mathrm{tr}\mathbf{E}^{(m)})^2 + \mu
\mathrm{tr}[(\mathbf{E}^{(m)})^2], \quad \mathbf{E}^{(m)} =
\frac{1}{m} \left[ (\mathbf{F}^T \mathbf{F})^{m/2} -\mathbf{I}
\right],
\end{aligned}
\end{equation}
where $\lambda$ and $\mu$ are the Lame constants and $m$ is a
parameter running over the reals. The case $m=0$ corresponds to Hencky
material, and $m=2$ corresponds to Saint Venant-Kirchhoff material.
The nominal stress is given by
\begin{equation}
\label{eq57}
\begin{aligned}
& \mathbf{S}=\mathbf{F}^{-1} \left[ \lambda
\frac{\mathrm{tr}(\mathbf{V}^{m}) -3}{m} \mathbf{V}^m + 2 \mu
\frac{\mathbf{V}^{2m} - \mathbf{V}^m}{m}\right],\quad
\mathbf{V}=\sqrt{\mathbf{F} \mathbf{F}^T}.
\end{aligned}
\end{equation}
After some simple algebra, we obtain the following expressions of the
two components of the nominal stress ($S_{Yy}$ is not needed)
\begin{equation}
\label{eq57_1}
\begin{aligned}
&S_{X\theta} = \frac{2\mu (\alpha r)^{m-1}}{1-2\nu} \left[(1-\nu)
\frac{(\alpha r)^m -1}{m} + \nu \frac{(r')^m -1}{m}
\right],\\
& S_{Z r} = \frac{2\mu (r')^{m-1}}{1-2\nu} \left[(1-\nu) \frac{(r')^m
-1}{m} + \nu \frac{(\alpha r)^m -1}{m} \right],
\end{aligned}
\end{equation}
where the parameter $\lambda$ has been replaced by Poisson's ratio
$\nu$ by $\lambda=2\mu \nu/(1-2\nu)$.

For the plate theory, we follow the previous procedure in
$(\ref{eq4})$ to expand $r$ in terms of $Z$
\begin{equation}
\label{eq58}
\begin{aligned}
& r(Z) = r_0 + r_1 Z + \frac{1}{2} r_2 Z^2+ \frac{1}{6} r_3 Z^3 +
\frac{1}{24} r_4 Z^4 + \cdots,\quad 0\leq Z \leq 2 h,
\end{aligned}
\end{equation}
where $r_i$ ($i=0,..,4$) are the constant coefficients. Accordingly as
in $(\ref{eq7},\ref{eq9})$, the deformation gradient and nominal
stress are expanded in series of $Z$. As we mentioned earlier, the
components $\mathbf{S}^{(i)}$ ($i=0,1,2$) are computed directly from
specific form $(\ref{eq57_1})$ by substituting $(\ref{eq58})$ rather
than using relations $(\ref{eq10})$. The explicit expressions are
given in Appendix B, and the component $\mathbf{S}^{(i)}$ ($i=0,1,2$)
depends linearly on the constant $r_{i+1}$. Then, by $(\ref{eq12})$
with $n=0,1$, the constants $r_i$ ($i=2,3$) can be expressed by $r_0$
and $ r_1$. For example
\begin{equation}
\label{eq59}
\begin{aligned}
&r_2 =\frac{- \alpha \bar{r}_0^{m-1} r_1^{2-m}
\left[\bar{r}_0^m-\left( \bar{r}_0^m+(m-1) r_1^m\right) \nu
-1\right]}{\nu \bar{r}_0^m - r_1^m (\nu -1)+m \left(-\nu \bar{r}_0^m+2
r_1^m (\nu -1)+1\right)-1},
\end{aligned}
\end{equation}
where $\bar{r}_0=\alpha r_0$. The expression of $r_3$ is given in
Appendix B. From the traction condition at $Z=0$ as in $(\ref{eqX1})$,
we obtain
\begin{equation}
\label{eq60}
\begin{aligned}
&r_1 = \left(\frac{\bar{r}_0^m \nu -1}{\nu -1}\right)^{\frac{1}{m}}.
\end{aligned}
\end{equation}
Finally, the plate equation $(\ref{eq17})_1$ furnishes an algebraic
equation for $\bar{r}_0$ upon using all the recursion relations
\begin{equation}
\label{eq61}
\begin{aligned}
&-3 \bar{r}_0^2 \left(\bar{r}_0^m-1\right) -\frac{3 \bar{h} \bar{r}_0
}{\nu -1}  \left(\frac{\bar{r}_0^m \nu -1}{\nu
-1}\right)^{\frac{1}{m}-1} [(2 \nu -m (\nu +2)+1) \bar{r}_0^m \\
&+2 (m-1) \nu  \bar{r}_0^{2 m}+m-1] + \bar{h}^2 f_1(\bar{r}_0) =0,
\end{aligned}
\end{equation}
where $\bar{h}=\alpha h$ and the explicit expression of $f_1$ is
omitted. Therefore, the quantity $\bar{r}_0$ is a function of the
parameter $\bar{h}$ for a fixed $m$. The algebraic equation
$(\ref{eq61})$ can be solved numerically without difficulty for any
fixed $m$, and for some special values of $m$ analytical solutions are
available.

The exact solutions in \cite{bruhns2002,xiao2011} for the pure bending
problem were obtained in integral form , and for some special values
of $m$, the integral solution can be reduced to explicit expressions
with elementary functions. We can compare the results for any $m$, but
in order to see clearly the error we only compare for $m=1$, for which
the exact solution and that of our plate equation are both explicit.
In our notations the exact solution in \cite{xiao2011} takes the form
\begin{equation}
\label{eq62}
\begin{aligned}
r=r^\star(Z)=-\frac{e^{\alpha Z} (1-2 \nu )+e^{2 \bar{h} }-e^{2
\bar{h}-\alpha Z}+1}{\alpha \left(1+e^{2 \bar{h}}\right) (\nu
-1)},\quad 0\leq Z \leq 2 h.
\end{aligned}
\end{equation}
For small thickness ratio $h$, one can do a Taylor expansion. Denoting
the corresponding coefficients by $r_i^\star (i=0,1,2,3)$, we have
\begin{equation}
\label{eq63}
\begin{aligned}
\bar{r}_0^\star =  &\alpha r_0^\star= 1 -\bar{h} + \frac{ \bar{h}^3}{3
}+O(\bar{h}^4),\quad r_1^\star = 1-\frac{\nu \bar{h}}{\nu -1} + \frac{
\nu \bar{h}^3}{3 (\nu -1)}+O(\bar{h}^4),\\
\frac{r_2^\star}{\alpha} = & \frac{\nu }{\nu -1} -\bar{h} + \frac{
\bar{h}^3}{3 } + O(\bar{h}^4), \quad \frac{r_3^\star}{\alpha^2}=  1
-\frac{\nu \bar{h}}{\nu-1}+ \frac{ \nu \bar{h}^3}{3 (\nu
-1)}+O(\bar{h}^4).
\end{aligned}
\end{equation}

In the plate theory, for $m=1$ the recursion relations for the
coefficients in $(\ref{eq58})$ become
\begin{equation}
\label{eq64}
\begin{aligned}
r_1 &= \frac{\bar{r}_0 \nu -1}{\nu -1},\quad r_2 =-\frac{\alpha (-\nu
\bar{r}_0 + \bar{r}_0-1)}{\nu -1},\quad r_3 =\alpha^2 r_1.
\end{aligned}
\end{equation}
These recursion relations are exact as we mentioned earlier. Indeed,
one can check easily that the Taylor expansion coefficients of the
exact solution $(\ref{eq62})$ also satisfy these relations. In
$(\ref{eq61})$, by setting $m=1$ the equation for $\bar{r}_0$ becomes
a linear equation, which immediately leads to the solution
\begin{equation}
\label{eq65}
\begin{aligned}
&\bar{r}_0 =\frac{3 -3 \bar{h}+ 2 \bar{h}^2}{3+ 2 \bar{h}^2 }.
\end{aligned}
\end{equation}
By further using Taylor expansions in $\bar{h}$, we obtain the first
four coefficients in the series of $r$
\begin{equation}
\label{eq66}
\begin{aligned}
\bar{r}_0 =& 1 -\bar{h} + \frac{2 \bar{h}^3}{3 }+O(\bar{h}^4),\quad
r_1 = 1-\frac{\nu \bar{h}}{\nu -1} + \frac{ 2 \nu
\bar{h}^3}{3 (\nu -1)}+O(\bar{h}^4),\\
\frac{r_2}{\alpha} = & \frac{\nu }{\nu -1} -\bar{h} + \frac{ 2
\bar{h}^3}{3 } + O(\bar{h}^4), \quad \frac{r_3}{\alpha^2}= 1
-\frac{\nu \bar{h}}{\nu -1 }+ \frac{ 2 \nu \bar{h}^3}{3 (\nu
-1)}+O(\bar{h}^4).
\end{aligned}
\end{equation}
Comparing $(\ref{eq63})$ and $(\ref{eq66})$, one can see that the
plate theory yields $O(\bar{h}^2)$-correct results for $r_i$ ($i=0,
1,2,3$) (the relative error is of $O(\bar{h}^3)$). As a result, the
current position $r(Z)$, strains and stresses are all pointwise
correct  up to $O(\bar{h}^2)$ for all bending angles. This example
gives a justification of our plate theory. As far as the authors are
aware of, no other plate theory has been shown that it can produce
results with a relative error of $O(h^3)$ for all relevant physical
quantities (displacements, strains and stresses) in comparison with
the exact solution.

By taking the parameter values $\alpha=\pi,\nu=0.3,h=0.1$, Figure
\ref{fig1b} shows the comparison between the plate solution and the
exact solution, with an almost indiscernible difference. Actually the
maximum relative error is less than $1.5\%$.  When $h$ increases to
$0.15$, the relative error is $5.1\%$.

\remark If we use the 2-D plate strain energy function in
$(\ref{eq48})$ to deal with this problem, the solution is found to be
\begin{equation}
\label{eq67}
\begin{aligned}
\bar{r}_0 =& \frac{3 (\nu -1) +3 (\nu +1) \bar{h}+ 2 (\nu -4)
\bar{h}^2}{3 (\nu -1) +6 \nu \bar{h}+8 (\nu -1) \bar{h}^2} =1 -\bar{h}
+ \frac{8 \bar{h}^3}{3 }+O(\bar{h}^4).
\end{aligned}
\end{equation}
Therefore, for the pure bending problem this approach also gives us an
$O(h^2)$- correct solution when the bending angle is prescribed,
although the higher-order $O(h^3)$ terms are different from
$(\ref{eq66})_1$. Comparing with the above results in $(\ref{eq66})$,
the maximum relative error in this approach  is increased by about $7$
times. Indeed, in numerical values this energy approach yields
relative errors of $10.5\%$ and $36.2\%$ respectively when $h=0.1$ and
$h=0.15$.

When the applied bending moment is prescribed, the consistent plate
theory can also provide $O(h^2)$-correct results as the plate boundary
condition $(\ref{eq21})$ agrees with this value up to the required
order. However, for the energy approach the generalized bending moment
(see $(\ref{eq53})_1)$ has to be assigned this prescribed value. But,
it does not conform with the bending moment in the exact formulation,
which further induces an $O(h^2)$ error.

\section{Concluding Remarks}

In this paper, a finite-strain plate theory, which is term-wise
consistent with the 3-D stationary potential energy principle, is
developed with no special restrictions on loadings or the order of
deformations. It seems that existing plate theories may not enjoy such
kind of consistency. The success relies on the derivation of the
recursion relations for the coefficients in the series expansion.
Another key is to use an expansion for the position vector about the
bottom surface, which enables the elimination of the second
coefficient exactly. Weak formulations suitable for finite element
calculations together with boundary conditions for some practical
cases are also provided. Besides the consistency, nice features of
this plate theory include: (a) reserving the local force-balance
structure in all three directions up to $O(h^2)$ (in a
through-thickness average); (b) obeying a 2-D virtual work principle; (c) not involving unphysical quantities like
higher-order stress resultants (usually present in existing
higher-order plate theories); (d) when the 3-D edge conditions are
known, no need to introduce the so-called generalized traction and
generalized bending moment (which do not conform with those in a 3-D
formulation); (e) providing up to $O(h^2)$-correct solution for the
pure bending problem of a rectangular block while other plate theories
have not been shown to give $O(h^2)$-correct results in comparison
with an exact solution.  In some textbook (e.g., \cite{ventsel2001}),
it was asserted that when the thickness ratio is bigger than some
value from $\frac{1}{10}$ to $\frac{1}{8}$ plate theories do not apply
any more and the 3-D theory should be used. However, for the pure
bending problem, our plate theory can actually provide accurate
results even when the thickness ratio reaches $0.15$, which breaks the
limit in the textbook. Also, in the widely used first-order
(Mindlin-Reissner) and third-order shear deformable plate theories for
small deflections, there are five differential equations (see
\cite{reddy2011}), while the present plate theory for finite strains
only has three differential equations. So, it appears that the present
plate theory is superior to the existing ones.

For the above-mentioned consistency and nice features, a price to pay
is that the plate theory cannot be formulated as a 2-D energy
minimization problem. By relaxing the consistency a little bit,  a 2-D
plate strain energy function is obtained by using those recursion
relations. Then, a plate problem can be solved through an energy
minimization. However, in terms of numerical errors, the consistent
plate theory produces a better result than that from this energy
approach for the pure bending problem. Another problem in this
approach is that the traction and bending moment conditions do not
conform with those in the 3-D formulation. We would also like to point
out that any plate theory involving generalized traction and
generalized bending moment has the same problem. Since they are not
equal to the average edge traction and bending moment in the 3-D
formulation, one would not expect that an $O(h^2)$-correct result can
be produced. Actually, one can see from $(\ref{eq40})$ that the
difference is $O(h^2)$, and such a difference will lead to an error of
$O(h^2)$ in the solution.

These two types of plate theories raise a philosophical question. Suppose that one plate theory provides an approximation to the governing system satisfied by the minimizer of the 3-D energy functional (like the first theory in this paper) and another plate theory provides a 2-D energy functional which is an approximation to the 3-D energy functional (like the second plate theory). The question is whether the 2-D minimizer in the latter approach approximates the 3-D minimizer to the required order. For example, if one substitute (1.1) into the traction conditions on the top and bottom surfaces, two vector equations for the coefficients are obtained (which the 3-D minimizer should satisfy). Whether the set of equations obtained from the 2-D energy variation (or virtual work principle) for the same coefficients are comparable to these two vector equations (to the required order) is not clear. Based on this reason, a plate theory which directly approximates the governing system for the 3-D minimizer seems to be preferred. Most existing plate theories belong to the 2-D energy approach (in principle), and this might be a reason why so far non plate theories have been shown to achieve $O(h^2)$-correct (or even $O(h)$-correct) results for all physical quantities.

It is known that so far the method of Gamma convergence fails to
provide higher-order (say, $O(h^2)$) plate theories. Since the present
consistent plate theory cannot be formulated as an energy minimization
problem, it might hint that the Gamma convergence could not succeed in
doing so as it is based on the energy minimization. For direct plate
theories, the difficulty is to propose the proper 2-D constitutive
relation. The derivations given in this paper indicate that such a
relation may not just be the property of the material and plate
geometry (see the discussions below $(\ref{eq36})$ and $(\ref{eq48})$
respectively) and depends on the tractions on the bottom/top surfaces
as well, which further increases the level of difficulty.

In our derivations, the body force is neglected for simplicity. But,
it is a straightforward matter to take it into account. For a dynamic
process, the 3-D field equations change to $\mathrm{Div} \mathbf{S}
=\rho \ddot{\mathbf{x}}$ with density $\rho$. Correspondingly, the 2-D
plate equations become $\nabla \cdot\bar{\mathbf{S}} =\rho
\ddot{\bar{\mathbf{x}}} -\bar{\mathbf{q}}$, with some modifications
for the intrinsic relations between $\mathbf{x}^{(k)}$ ($k=1,2,3,4$)
and $\mathbf{x}^{(0)}$. We shall represent the details elsewhere,
together with a study on the degenerated linear plate theory. This
work is currently being extended for the derivation of a consistent
shell theory.

Finally, we would like to mention that under the imposed smooth
assumptions (see, e.g. $(\ref{eq4})$), the plate theory agrees with
the 3-D weak formulation to $O(h^3)$ rigorously  and the order of
errors in the plate equations and boundary conditions is also
rigorous. However, the proof that those errors will lead to an error
of the same order for the solution is not provided, which will be left
for a future investigation. Plate theories can be dated back at least
150 years ago, and it was stated by Steigmann in \cite{steigmann2007}:
"The derivation of two-dimensional theories of elastic plates from
three-dimensional elasticity is one of the main open problems in solid
mechanics".  In terms of the formal procedure, the agreement with the
3-D weak formulation and the order of error in the field equations and
boundary conditions, the present work makes a contribution toward to the
resolution of this open problem. Desirably, comparisons between
solutions of the present plate theory and those in the 3-D formulation
in more numerical examples are needed to see whether this plate theory 
provides a plausible solution, which will be carried out in the future.

\section*{Acknowledgment}

The authors would like to thank professor P. G. Ciarlet for providing
some relevant references.

\section*{Appendix}

\appendix

\numberwithin{equation}{section}

\section{The differential system from the 2-D plate strain energy}

By taking variation of the 2-D plate potential energy in
$(\ref{eq48})$ and setting it to be zero, we can obtain
\begin{equation}
\label{eq51}
\begin{aligned}
&\nabla \cdot \hat{\mathbf{S}}  =- \bar{\mathbf{q}},\quad in \quad \Omega,\\
&\mathbf{x}^{(0)}= \mathbf{b}^{(0)}(s),\quad \mathbf{x}^{(1)}=
\mathbf{b}^{(1)}(s),\quad on \quad \partial\Omega_0,\\
&\hat{\mathbf{S}} ^T \mathbf{N} =\mathbf{q}_0, \quad
\tilde{\mathbf{S}}^T \mathbf{N}= \mathbf{q}_1,\quad on \quad
\partial\Omega_q.
\end{aligned}
\end{equation}
where the two stress tensors $\hat{\mathbf{S}}$ and
$\tilde{\mathbf{S}}$ are expressed as
\begin{equation}
\label{eq52}
\begin{aligned}
&\hat{\mathbf{S}}=\mathbf{S}^{(0)} + h  \mathbf{S}^{(1)} + \frac{2}{3}
h^2 \hat{\mathbf{S}}^{(2)}  + 2h \mathbf{P}_2,\quad
\tilde{\mathbf{S}}= \mathbf{S}^{(0)} + \frac{4h}{3}
\mathbf{S}^{(1)} + 2h \mathbf{P}_1,\\
& \hat{\mathbf{S}}^{(2)}= {\mathcal{A}}^2 [\mathbf{F}^{(1)} ,
\mathbf{F}^{(1)}] - \mathbf{x}^{(2)}\cdot (\nabla \cdot
{\mathcal{A}}^1),\\
& \mathbf{p}_0=\nabla \cdot \mathbf{S}^{(0)} + \bar{\mathbf{q}},\quad
\mathbf{p}_1 = \mathbf{p}_0 + \frac{2}{3}h \nabla \cdot
\mathbf{S}^{(1)} - \frac{1}{3} h \hat{\mathbf{S}}^{(2)T}
\mathbf{k},\\
& \mathbf{P}_0 =  \frac{\partial \mathbf{B}^{-1}}{\partial
\mathbf{F}^{(0)}} [\mathbf{p}_0 \otimes \mathbf{f}^{(2)}] +
{\mathcal{A}}^2 [\nabla \mathbf{x}^{(1)}, \mathbf{B}^{-1} \mathbf{p}_0
\otimes \mathbf{k}] + \mathbf{B}^{-1} \mathbf{p}_0 \cdot
(\nabla \cdot {\mathcal{A}}^1),\\
& \mathbf{P}_1 = {\mathcal{A}}^1 [\mathbf{B}^{-1} \mathbf{p}_0 \otimes
\mathbf{k}],\quad \mathbf{P}_2= {\mathcal{A}}^1 [\mathbf{B}^{-1}
(\mathbf{p}_1 + h \nabla \cdot \mathbf{P}_1 -h \mathbf{P}_0^T
\mathbf{k}) \otimes \mathbf{k} ] + h \mathbf{P}_0.
\end{aligned}
\end{equation}
The boundary conditions in $(\ref{eq51})$ are parallel to those of
cases 1 and 2 of subsection 3(c).

Alternatively, if the variation of the load potential the 2-D energy
$\delta \bar{V}_{2D}$ in $(\ref{eq49})$ is used, then the boundary
conditions parallel to those in cases 3 to 6 of section 4 will come
out. Specifically, the quantities $\bar{\mathbf{S}}$ and $\mathbf{M}$
in $(\ref{eq38}-\ref{eq42})$ are replaced by the newly defined
averaged stress $\hat{\mathbf{S}}$ and newly defined moment tensor
$\mathbf{\hat{M}}$
\begin{equation}
\label{eq53}
\begin{aligned}
& \mathbf{\hat{M} [N,N]} = h \mathbf{B}_1^T \mathbf{B}^{-1}
\tilde{\mathbf{S}}^T \mathbf{N},\quad \mathbf{\hat{M} [N,T]} =h
\mathbf{B}_3^T \mathbf{B}^{-1} \tilde{\mathbf{S}}^T \mathbf{N},
\end{aligned}
\end{equation}
where $\mathbf{\hat{M} [N,N]}$ is considered as a generalized bending
moment.

Note that the explicit expressions for $\hat{\mathbf{S}}$,
$\tilde{\mathbf{S}}$ and $\mathbf{\hat{M}}$ in $(\ref{eq52},
\ref{eq53})$ are only for comparison with the consistent plate system
in section 3. In practice they can be derived directly from the
explicit form of $\Phi_{2D}$ in $(\ref{eq48})$. Substituting the
relation $(\ref{eq14})$ for $\mathbf{x}^{(2)}$, one can write
$\Phi_{2D}=\Phi_{2D}(\nabla \mathbf{x}^{(0)},\mathbf{x}^{(1)},\nabla
\mathbf{x}^{(1)})$. Then $\tilde{\mathbf{S}}$ is calculated by
\begin{equation}
\label{eq53_3}
\begin{aligned}
\tilde{\mathbf{S}} = \frac{1}{2h^2}\frac{\partial \Phi_{2D}}{ \partial
(\nabla \mathbf{x}^{(1)})}.
\end{aligned}
\end{equation}
Further substituting the relation for $\mathbf{x}^{(1)}$, one can
write $\Phi_{2D}=\Phi_{2D}(\nabla \mathbf{x}^{(0)},\nabla \nabla
\mathbf{x}^{(0)})$. Then $\hat{\mathbf{S}}$ and $\mathbf{\hat{M}}$ are
given by
\begin{equation}
\label{eq53_4}
\begin{aligned}
\mathbf{M} = \frac{1}{2h} \frac{\partial \Phi_{2D}}{\partial
(\nabla\nabla\mathbf{x}^{(0)})},\quad \hat{\mathbf{S}} = \frac{1}{2h}
\frac{\partial \Phi_{2D}}{\partial (\nabla\mathbf{x}^{(0)})} -
\nabla\cdot\mathbf{M}.
\end{aligned}
\end{equation}

Comparing with $(\ref{eq17})$, we see that the present plate equation
$(\ref{eq51})_1$ does not reserve the force-balance structure
inherited from the 3-D system in all three directions. However, as
mentioned earlier, with a 2-D plate strain energy function, the plate
problem can be considered as an energy minimization problem, which
could have advantages in some problems.

\section{Some expressions in the pure-bending problem}

The components in the series of the nominal stress are given by
\begin{equation}
\label{eq53_1}
\begin{aligned}
S_{X \theta}^{(0)}= &\frac{2 \mu  \bar{r}_0^{m-1} \left(-\nu
r_1^m+(\nu -1) \bar{r}_0^m+1\right)}{m (2 \nu -1)},\quad
S_{Zr}^{(0)}=\frac{2 r_1^{m-1} \mu  \left(-(\nu -1)
r_1^m+\nu \bar{r}_0^m-1\right)}{m-2 m \nu },\\
S_{X \theta}^{(1)}=& -\frac{2 \mu  \bar{r}_0^{m-2} \left(m r_2 \nu
\bar{r}_0 r_1^m-\alpha (2 m-1) (\nu -1) \bar{r}_0^m r_1^2+\alpha (m-1)
\left(r_1^m \nu -1\right) r_1^2\right)}{m r_1 (2 \nu -1)},\\
S_{Zr}^{(1)}=& -\frac{2 r_1^{m-2} \mu  }{m (2 \nu -1)
\bar{r}_0}\left[\alpha m r_1^2 \nu \bar{r}_0^m+(m-1) r_2 \nu
\bar{r}_0^{m+1}\right.\\
&\left.-r_2 \left(-(\nu -1) r_1^m+m \left(2 (\nu -1)
r_1^m+1\right)-1\right) \bar{r}_0\right],\\
S_{X \theta}^{(2)}=& \frac{2 \mu  \bar{r}_0^{m-3} }{m r_1^2 (2 \nu
-1)}\left[-m \left((m-1) r_2^2+r_1 r_3\right) \nu  \bar{r}_0^2
r_1^m\right.\\
&\left.+2 \alpha^2 \left(2 m^2-3 m+1\right) (\nu -1) \bar{r}_0^m r_1^4
-\alpha^2 \left(m^2-3 m+2\right) \left(r_1^m \nu -1\right)
r_1^4\right.\\
&\left. +\alpha (2 m-1) r_2 (\nu -1) \bar{r}_0^{m+1} r_1^2-\alpha
(m-1) r_2 \left((2 m+1) r_1^m \nu -1\right) \bar{r}_0
r_1^2\right],\\
S_{Zr}^{(2)}= & \frac{2r_1^{m-3} \mu  }{m (2 \nu -1)
\bar{r}_0^2}\left[-\alpha^2 (m-1) m r_1^4 \nu \bar{r}_0^m-\alpha m (2
m-1) r_1^2 r_2 \nu \bar{r}_0^{m+1}\right.\\
& -(m-1) \left((m-2) r_2^2+r_1 r_3\right) \nu \bar{r}_0^{m+2}+((m-1)
(-2 (\nu -1) r_1^m  +m r_2^2 \\
&\left(4 (\nu -1) r_1^m+1\right)-2)  +r_1 r_3 \left(-(\nu -1) r_1^m+m
\left(2 (\nu -1) r_1^m+1\right)-1\right)) \left. \bar{r}_0^2\right].
\end{aligned}
\end{equation}
The recursion relation for $r_3$ is given by
\begin{equation}
\label{eq53_2}
\begin{aligned}
r_3= & \frac{r_1^{-m-1} }{\bar{r}_0^2 \left((2 m-1) (\nu -1)
r_1^m-(m-1) \nu  \bar{r}_0^m+m-1\right)}\left[-(m-1) r_2^2 \right.\\
&\left(2 (2 m-1) (\nu -1) r_1^m+m-2\right) \bar{r}_0^2 r_1^m+\alpha^2
(2 m-1) (\nu -1) \bar{r}_0^{2 m} r_1^4\\
&\left.+(m-1) \bar{r}_0^m \left(r_2 \nu \bar{r}_0 \left(2 \alpha m
r_1^2+(m-2) r_2 \bar{r}_0\right) r_1^m+\alpha^2 \left((m-1) \nu
r_1^m+1\right) r_1^4\right)\right].
\end{aligned}
\end{equation}

\end{document}